\documentclass{aa}
\usepackage{graphicx}
\usepackage{txfonts}
\usepackage{subcaption}
\usepackage{float}
\usepackage{enumitem}
\authorrunning{I.C. Jebaraj, et al.,}
\titlerunning{Radio triangulation of type II bursts}
\begin{document} 

   \title{Using radio triangulation to understand the origin of two subsequent type II radio bursts}

   \author{I. C. Jebaraj
          \inst{1,2},
          J. Magdaleni\'{c}\inst{1},
          T. Podladchikova\inst{3},
          C. Scolini\inst{2,1},
          J. Pomoell\inst{4},
          A. M. Veronig\inst{5,6},
          K. Dissauer\inst{5},\\
          V. Krupar\inst{7,8,9},
          E. K. J. Kilpua\inst{4},
          and S. Poedts\inst{2,10}
          }

   \institute{1. SIDC, Royal Observatory of Belgium, Brussels, Belgium\\
             2. Centre for mathematical Plasma Astrophysics, KU Leuven, Leuven, Belgium\\
             3. Skolkovo Institute of Science and Technology, Moscow, Russia\\
             4. University of Helsinki, Helsinki, Finland\\
             5. Institute of Physics, University of Graz, Graz, Austria\\
             6. Kanzelh\"ohe Observatory for Solar and Environmental Research, University of Graz, Austria\\
             7. Universities Space Research Association, Columbia, MD 21046, USA\\
             8. NASA Goddard Space Flight Center, Greenbelt, MD 20771, USA\\
             9. Institute of Atmospheric Physics of the Czech Academy of Sciences, Prague 14131, Czech Republic\\
             10. Institute of Physics, University of Maria Curie-Sk{\l}odowska, PL-20-031 Lublin, Poland}

%   \date{Received September 15, 1996; accepted March 16, 1997}

% \abstract{}{}{}{}{} 
% 5 {} token are mandatory

    \abstract
   {Eruptive events such as coronal mass ejections (CMEs) and flares accelerate particles and generate shock waves which can arrive at Earth and can disturb the magnetosphere. Understanding the association between CMEs and CME-driven shocks is therefore of high importance for space weather studies.}
   {We present a study of the CME/flare event associated with two type II bursts observed on September 27, 2012. The aim of the study is to understand the relationship between observed CME and the two distinct shock wave signatures.}
   {The multiwavelength study of the CME/flare event was complemented with radio triangulation of the associated radio emission and modeling of the CME and the shock wave employing MHD simulations.}
   {We found that, although temporal association between the type II bursts and the CME is good, the low frequency type II (LF-type II) burst occurs significantly higher in the corona than the CME and its relationship to the CME is not straightforward.
   The study of the EIT wave (coronal bright front) shows the fastest wave component to be in the south-east quadrant of the Sun. This is also the quadrant in which the source positions of the LF-type II were found to be located, probably resulting from the interaction between the shock wave and a streamer.}
   {The relationship of the CME/flare event and shock wave signatures are discussed using the temporal association, as well as the spatial information of the radio emission. Further, we discuss the importance and possible effects of frequently non-radial propagation of the shock wave.}

   \keywords{Sun: radio radiation --
                Sun: particle emission --
                Sun: heliosphere --
                Sun: coronal mass ejections (CMEs) --
                magnetohydrodynamics (MHD) --
                shock waves
               }
               
\maketitle

%________________________________________________________________
\section{Introduction} \label{Sec:Introduction}

%%%%%%%%%%%%%%%%%%%%%%%%%%%%%%%%%%%%%%%%%%%%%%%%%%%%%%%%%%%%%%
 \begin{figure*}[h]
  \centering
  \begin{subfigure}[b]{0.99 \textwidth}
    \includegraphics[width=\textwidth]{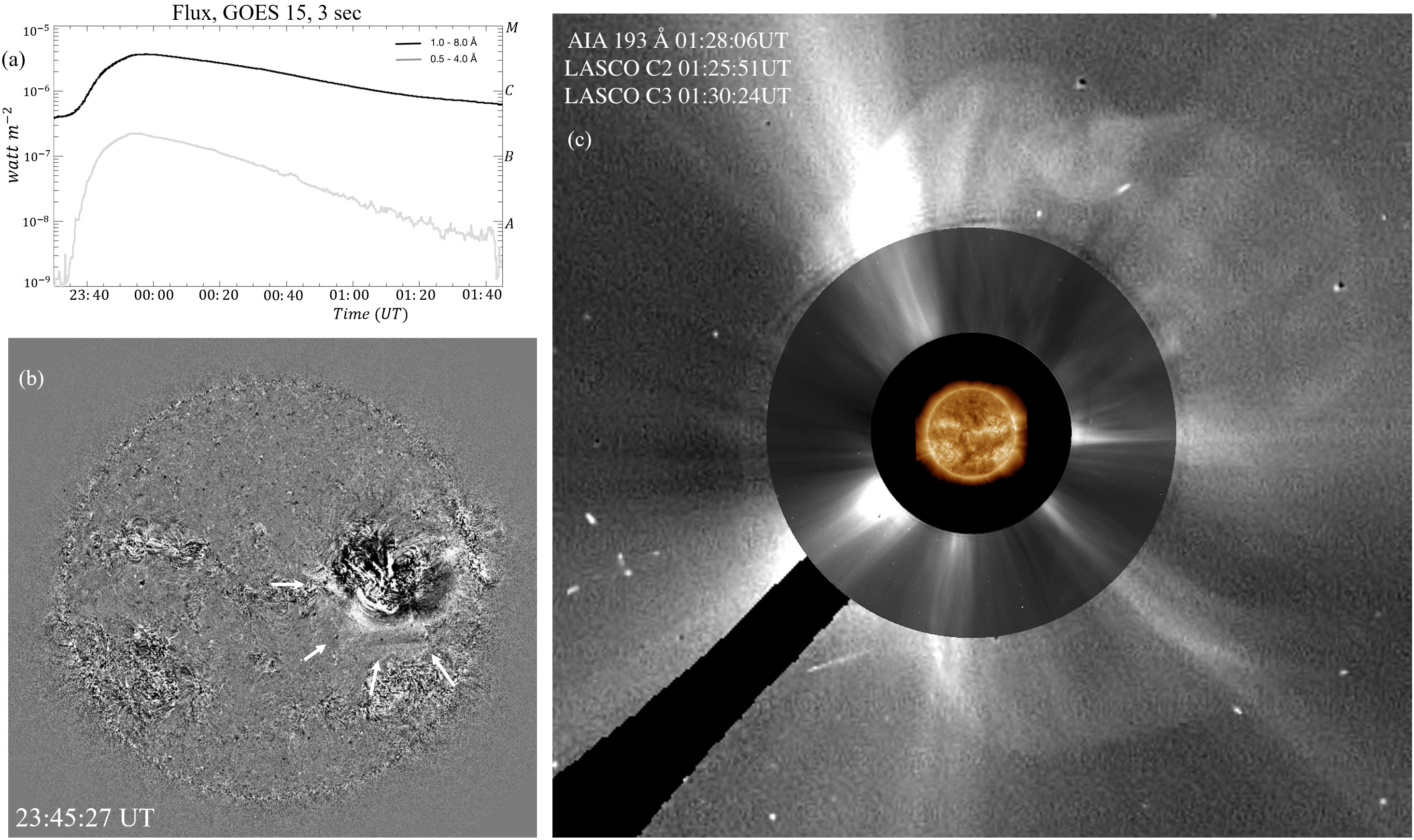}
  \end{subfigure}
  \caption{Overview of the CME/flare event: (a) The \textit{GOES} C3.7 X-ray flare curve shows a rather long flare decay phase. (b) Running difference image of the \textit{SDO/AIA} 193\AA~~channel. The white arrows mark what appears to be an EIT wave (video available in the online version of the article). (c) Combined images of the \textit{SDO/AIA} observations at 193\AA, \textit{SOHO/LASCO C2} and \textit{C3} observations around 01:30~UT.}
  \label{Fig:WL_Observations}
  \end{figure*}
%%%%%%%%%%%%%%%%%%%%%%%%%%%%%%%%%%%%%%%%%%%%%%%%%%%%%%%%%%%%%%%%

Large scale energy release in the solar corona can appear in the form of coronal mass ejections \citep[CMEs, e.g.,][]{Shibata11, Green18} and flares \citep[e.g.,][]{Fletcher11, Benz17}. During these eruptive phenomena, particles are accelerated \citep[][]{Miteva17}, plasma is heated and waves and shocks are generated \citep{Aschwanden19book}. The shock waves associated with eruptive events can manifest via a variety of signatures. Chromospheric Moreton waves, EIT waves \citep[coronal bright fronts associated with solar eruptions, see e.g.][and references therein]{Zhukov04}, and type II radio bursts are often considered to be signatures of the same shock wave propagating in the solar corona \citep[e.g.,][and references therein]{Warmuth04, Vrsnak06, Veronig06, Veronig10, Warmuth15}. We will focus on type II radio bursts which are the longest known signatures of shock waves in the solar corona \citep{Wild50b}, and are also excellent means for tracking the shock wave propagation \citep[e.g.,][]{Wild50a, Ginzburg58, Melrose80, Klassen99, Magdalenic12}. 

Type III radio bursts \citep[radio signatures of electron beams travelling along open and quasi-open magnetic field lines, see e.g.][]{Reid14Review} are also often observed in association with eruptive phenomena \citep[e.g.,][]{Reiner98, Reiner01, Cairns03, Cremades07, Reid14, Krupar15}. Both type II and type III bursts are generally considered to be plasma emissions, generated by beams of supra-thermal electrons. Type II and type III bursts are observed at both, the fundamental plasma frequency (\(f_{pe}\)) as well as the second harmonic (\(2f_{pe}\)). Sometimes only one of the two components is observed \citep[for review, see e.g.][and references therein]{Melrose17}. As the sources of radio emission propagate away from the Sun, radio emission occurs at progressively lower frequencies which corresponds to the decrease of the ambient electron density. Type II radio bursts observed at metric wavelengths (around 100~MHz) are generally considered to be signatures of shock waves propagating in the low solar corona while emission in the hectometric to kilometric wavelengths is associated with shock waves propagating through the outer corona to interplanetary space. 

Understanding the origin of coronal shock waves and associated type II emission is a complex, and widely discussed problem \citep[e.g.][]{Gary84, Klein99, Maia00, Magdalenic08, Magdalenic10, Nindos11, Zimovets12}. Distinguishing the shock driver, particularly in the low corona, is often a difficult task, mostly due to the good temporal synchronisation between the flare impulsive phase and the acceleration phase of the CME. Although some shocks appear to be generated by flares \citep[e.g.,][]{Magdalenic10, Magdalenic14, KumarP16, Eselevich19}, the majority of shock waves are CME-driven \citep[e.g.,][and references therein]{VrsnakC08}. Even when the type II emission is clearly a signature of the CME-driven shock wave, the relative position of the type II sources and the shock driver is unclear, i.e. whether the emission originates from close to the CME flank or the CME nose. A number of studies have demonstrated that the radio emission is most probably originating from the regions close to the CME flank \citep[][]{Reiner98, Magdalenic12, Shen13, Magdalenic14, MartinezOliveros15, Krupar16, Krupar19}. Only occasionally events are reported with the type II emission situated close to the CME nose regions \citep[][]{MartinezOliveros12, Makela16, Makela18}. Similarly, it was shown that coronal EIT waves are initiated by the fast acceleration of the CME flanks \citep{Veronig08, Kienreich09, Patsourakos09, Long17, Veronig18}.

In this paper, we present a study of a CME/flare event on September 27/28, 2012 and the associated radio event. We investigate the complex relationship between the CME, the shock wave, and the origin of the two associated type II radio bursts.  We discuss the importance of the effects induced by the non-radial propagation of the CME-driven shock wave and the consequence on the associated radio emission. The observations employed in the study are introduced in Section \ref{Sec:Observations}, followed by the description of the CME/flare event (Section \ref{Sec:CME_Observations}) and it's propagation (Section \ref{Sec:CME_propagation}). The study of the EIT wave associated with the eruption  is presented in Section \ref{Sec:EIT_Wave}. The radio event is reported and is analysed employing the classical method in Section \ref{Sec:Radio_analysis} which is followed by the results of the radio triangulation study in Section \ref{Sec:Radio_triangulation}. 
An interpretation of the results with regards to the ambient coronal conditions is discussed in Section \ref{Sec:Coronal_conditions}. The study is briefly summarised and the effects of radio wave propagation are discussed in Section \ref{Sec:Summary}, and finally, the most important findings of the study are listed in Section \ref{Sec:Conclusions}, respectively.

\section{Observational data} \label{Sec:Observations}

The multi-wavelength study of the September 27/28, 2012 event employs  white light (WL), radio, extreme ultra violet (EUV), magnetogram, and X-ray observations.

\subsection{White light coronagraph observations} \label{Sec:WL_Observations}

We used coronagraph observations from different instruments and viewpoints: (a) The Large Angle and Spectroscopic Coronagraph \citep[\textit{LASCO};][]{Brueckner95} on board the Solar and Heliospheric Observatory \citep[\textit{SOHO};][]{Domingo95} mission provides two coronagraphs, C2 and C3, with different field of view, (b) and the Solar TErrestrial RElations Observatory Ahead and Behind \citep[\textit{STEREO A} \& \textit{STEREO B};][]{Kaiser08} coronagraphs \textit{COR 1} and \textit{COR 2} \citep{Howard08}.

\subsection{EUV, magnetogram, and X-ray observations} \label{Sec:EUV_Observations}

Observations of the Sun at the EUV wavelengths are often used in studies of the evolution of active regions, flares, waves, and on-disk signatures of CMEs. In this study we employed observations from: (a) the Extreme Ultra Violet Imagers \citep[\textit{EUVI}; ][]{Howard08} instrument on-board \textit{STEREO} which observe the solar corona with a cadence of 15 minutes in four EUV passbands. (b) The Atmospheric Imaging Assembly \citep[\textit{AIA};][]{Lemen12} onboard Solar Dynamics Observatory \citep[\textit{SDO};][]{Pesnell12} which routinely provides high-cadence, high-resolution EUV images of the Sun from Earth's orbit. 

Additionally, we also employ soft X-ray observations by the Geostationary Operational Environmental Satellite \citep[\textit{GOES} 15][]{Howard94}. 

%%%%%%%%%%%%%%%%%%%%%%%%%%%%%%%%%%%%%%%%%%%%%%%%%%%%%%%%%%%%
 \begin{figure}[ht]
   \centering
   \begin{subfigure}[b]{0.49 \textwidth}
    \includegraphics[width=\textwidth]{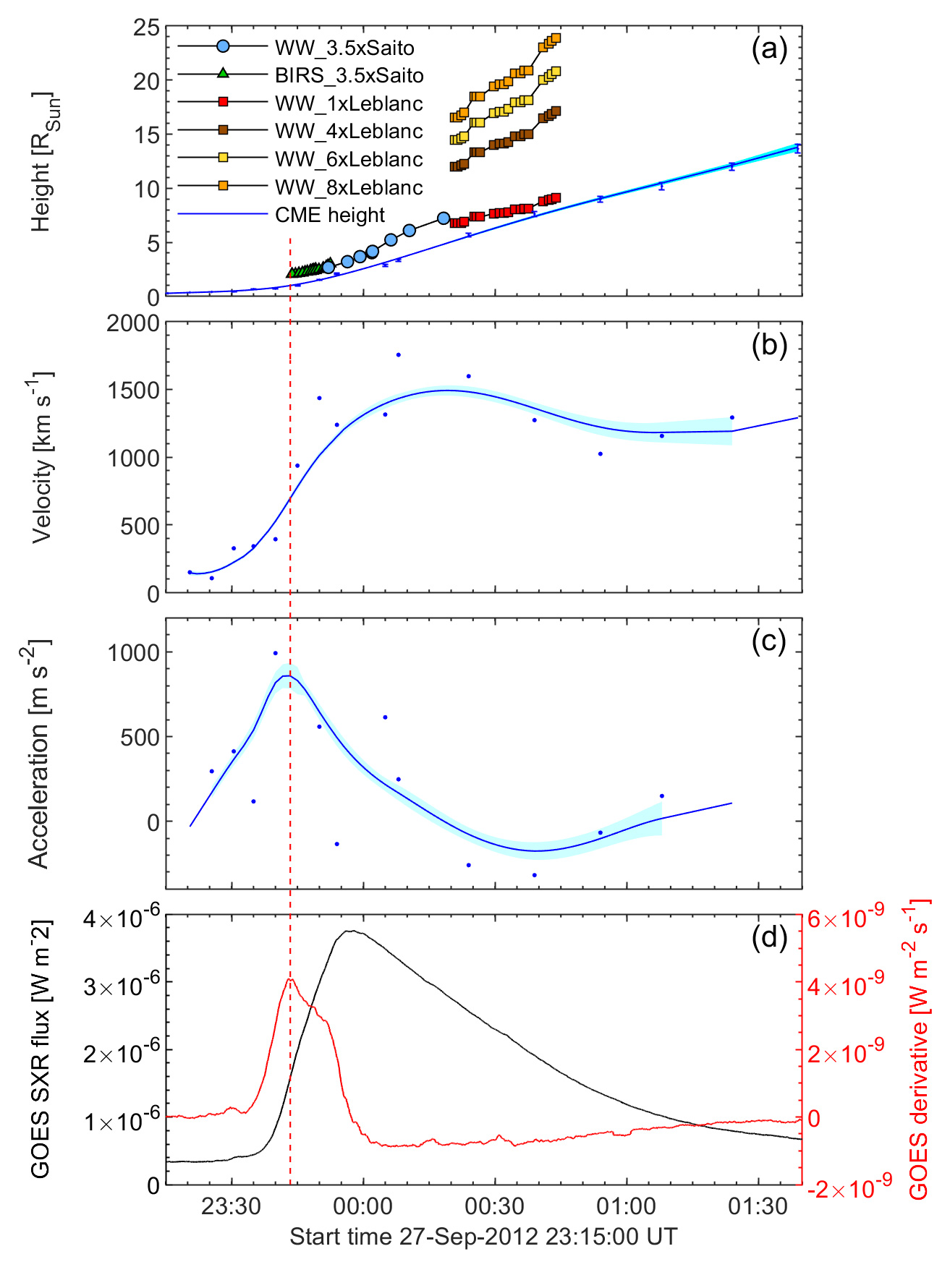}
   \end{subfigure}
   \caption{CME kinematics based on \textit{STEREO A/EUV} and coronagraph images. \textit{(a)} CME height (blue dots) together with error bars and radio emission heights. The corresponding line shows the smoothed height–time profile. HF-type II heights are obtained using a 3.5-fold Saito, while the LF-type II heights are obtained using 1-fold, 4-fold, 6-fold, and 8-fold Leblanc based on radio triangulation results discussed in Section \ref{Sec:Discussion}. \textit{(b)} CME velocity and \textit{(c)} acceleration profile obtained 
   by numerical differentiation of the data points (circles) and the smoothed curves (lines). The shaded areas represent the error ranges obtained from the smoothed curves. \textit{(d)} \textit{GOES} soft X-ray flux and it's derivative. The red vertical line denotes the peak of the \textit{GOES} derivative}
   \label{Fig:CME_Kinematics}
   \end{figure}
%%%%%%%%%%%%%%%%%%%%%%%%%%%%%%%%%%%%%%%%%%%%%%%%%%%%%%%%%%%%%%%% 

\subsection{Radio observations} \label{Sec:Radio_Observations}

In this study we used observations from the following ground based and space based observatories: (a) dynamic spectra from Bruny Island Radio Spectrometer \citep[BIRS; ][]{Erickson97}, covering the decametric range (80-20~MHz). (b) Dynamic spectra from Culgoora, which covers the metric and decametric range (1800-18~MHz). (c) Dynamic spectra from the \textit{STEREO/WAVES} instruments \citep[][]{Kaiser05, Kaiser08, Bougeret08} are routinely available in the frequency range, 2.5-16025~kHz and the high-frequency receiver (HFR) provides instantaneous direction-finding measurements at a number of discrete frequencies in the range 125-1975~kHz \citep{Cecconi08}. And (d) dynamic spectra from the \textit{Wind/WAVES} \citep[][]{Bougeret95} instrument is available in the frequency range, 4-13825~kHz and the RAD1 receiver provides direction-finding measurements (at selected frequencies; 100-1040~kHz). 

%which consists of three receivers, TNR, RAD1, and RAD2, providing dynamic spectra in the frequency ranges, 4-256~kHz, 20-1040~kHz and 1075-13825~kHz, respectively. RAD1 provides direction-finding measurements (at selected frequencies; 100-1040~kHz) that we used in this study. 

\section{Event description} \label{Sec:CME_Observations}

The \textit{GOES} C3.7 flare (23:35-23:47-1:40~UT) was associated with a two-step filament eruption and a full-halo CME first observed in the \textit{SOHO/LASCO C2} field of view at 00:00~UT on September 28, 2012 \citep[Fig.~\ref{Fig:WL_Observations}a and \ref{Fig:WL_Observations}c, respectively; studied in][]{Veronig19}. We also observed on-disk signatures of the CME in the form of a coronal dimming and an EIT wave. The CME/flare event originated from NOAA active region 11577 (N09,W31) having a $\beta \gamma$ configuration of its photospheric magnetic field at the time of eruption. The flare and the off-limb signatures of the CME were observed by \textit{SDO/AIA}, together with a well-defined EIT wave propagating mainly in the south-east direction from the active region (best observed in the at AIA 193~\AA~filter, Fig.~\ref{Fig:WL_Observations}b). The \textit{STEREO A/COR 2} and \textit{STEREO B/COR 2} coronagraphs observed the CME for the first time at 00:12~UT and 01:03~UT, respectively. The CME was also observed by both \textit{STEREO A} and \textit{STEREO B} Heliospheric imagers (\textit{HI}) and is included in the Heliospheric Cataloging, Analysis and Techniques Services catalogue (HELCATS; \url{https://www.helcats-fp7.eu/}). A WL shock wave observed by all three coronagraphs accompanied the CME under study. 

%%%%%%%%%%%%%%%%%%%%%%%%%%%%%%%%%%%%%%%%%%%%%%%%%%%%%%%%%%%%
\begin{figure*}[ht]
     \centering
     \begin{subfigure}[b]{0.99\textwidth}
         \centering
         \includegraphics[width=\textwidth]{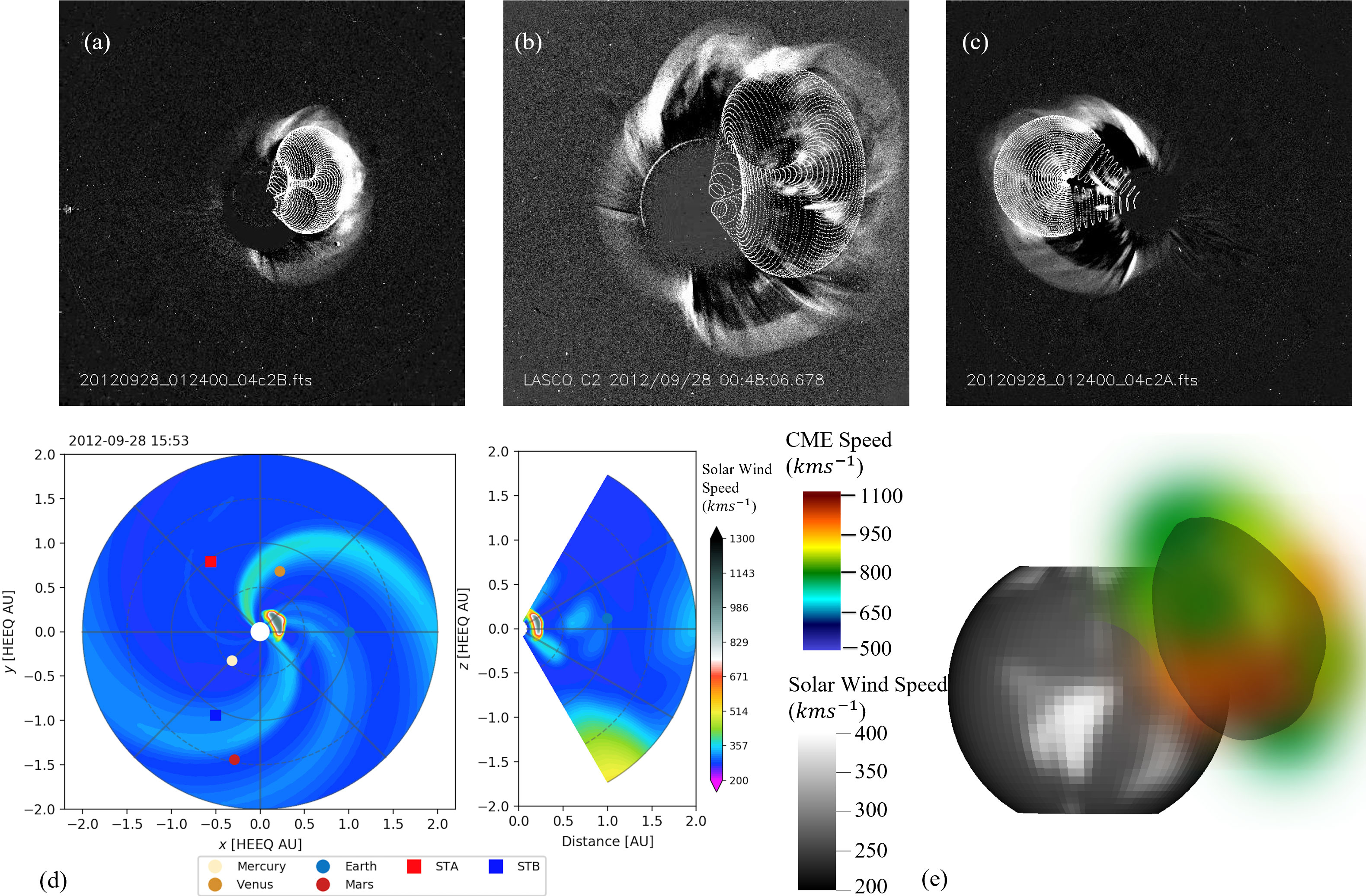}
         \label{}
     \end{subfigure}
     \hfill
     \caption{Panels \textit{(a), (b)}, and \textit{(c)} show the graduated cylindrical shell (GCS) reconstruction of the CME body.  Coronagraph images from: \textit{(a)} \textit{STEREO B/COR 2}, \textit{(b)} \textit{SOHO/LASCO C3}, and \textit{(c)} \textit{STEREO A/COR 2} at roughly the same time (September 28, 2012 at 01:24~UT) are shown.  
     Panels \textit{(d)}, and \textit{(e)} show CME modeled by EUHFORIA.
     \textit{(d)} Propagation of the modelled cone CME in the ecliptic and meridional perspectives. \textit{(e)} The CME speeds at the cone surface, as seen from Earth view. The grey sphere shows the solar wind radial velocity at the inner boundary of EUHFORIA (0.1 AU).} 
     \label{Fig:All_models}.
\end{figure*}

%%%%%%%%%%%%%%%%%%%%%%%%%%%%%%%%%%%%%%%%%%%%%%%%%%%%%%%%%%%%%%%%%%

In order to understand the possible preconditioning of the solar corona, i.e. the possible existence of large preceding eruptions which could have perturbed the global structure of the corona and influenced the propagation of the CME and associated shock \citep[e.g.][]{Liu14, Temmer15}, we investigate the solar events in a time window of 12~h previous to the studied event. The only CME/flare event (hereafter, event-0) possibly associated with the studied one was a back-sided halo-CME originating from NOAA AR 11574. Two subsequent eruptions were observed in the \textit{STEREO A/COR 1} field of view, starting shortly before 10:00~UT on September 27, and they were accompanied with intense radio event . The two \textit{STEREO} spacecraft observed the radio event across the entire frequency range, while \textit{Wind/WAVES} observations show only the low frequency part of the radio emission (the high frequency part was occulted by the Sun.) This back-side event could not have significantly influenced the development and the propagation of the main studied event. The particularity of the radio event-0 will be discussed in a separate publication.

\section{CME propagation} \label{Sec:CME_propagation}
In order to study the CME kinematics we use measurements of the CME height derived from \textit{STEREO A/EUVI}, \textit{COR 1}, and \textit{COR 2} images (Fig.~\ref{Fig:CME_Kinematics}a). \textit{STEREO A} had the best view to observe the CME evolution close to the limb, thus minimising projection effects on the derived kinematic profiles. Fig. \ref{Fig:CME_Kinematics} shows the CME kinematics along the position angle of ${65}^\circ$.  The CME velocity (Fig.~\ref{Fig:CME_Kinematics}b) and acceleration (Fig.~\ref{Fig:CME_Kinematics}c) profiles were obtained by smoothing the height-time data and deriving the first and second time derivatives \citep{Dissauer2019}. The smoothing algorithm that we use for approximating the curves \citep{Podladchikova2017}, was extended toward non-equidistant data. From the obtained acceleration profiles we interpolate to equidistant data points based on minimisation of the second derivatives, and reconstruct the corresponding velocity and height profiles by integration.
The projected speed of the CME leading edge reaches a peak value of 1490~km/s, and the CME acceleration peaks on September 27 at 23:43~UT with a value of 860 m/s$^2$.

\subsection{Modelling of the CME with EUHFORIA} \label{Sec:CME_modelling}

In order to understand the relationship between the CME and shock wave we modelled the CME using two complementary approaches, through forward modelling using the GCS model, and magnetohydrodynamics (MHD) modelling using EUHFORIA. (Fig.~\ref{Fig:All_models}).

We apply the graduated cylindrical model (GCS), a simple geometric reconstruction technique developed by \cite{Thernisien06, Thernisien09}, using coronagraph images from multiple viewpoints, i.e. \textit{SOHO/LASCO C2} and \textit{C3}, and \textit{STEREO A} \& \textit{B} \textit{COR 2}. 
This technique is based on fitting the observed white-light structure of the CME using a croissant-like three-dimensional shell which, when applied to a sequence of imaging observations, allows to determine the kinematic and geometric properties of the CME. These are then used as input for the heliospheric MHD simulations. 
Fig.~\ref{Fig:All_models}a, b, and c show the results of the reconstruction of the CME on September 28 around 01:24~UT, resulting in the following CME parameters (in Stonyhurst coordinates):
CME latitude $\theta_\mathrm{CME} = 20^\circ$, 
longitude $\phi_\mathrm{CME} = 30^\circ$, 
front height $h_\mathrm{CME} = 11.9$~$R_{\odot}$, 
aspect ratio $\kappa_\mathrm{CME} = 0.30$, 
leg angle $\alpha_\mathrm{CME} = 20^\circ$, 
and tilt $\gamma_\mathrm{CME} = -90^\circ$.
The estimated 3D speed of the CME is found to be 1270~km/s at 01:24~UT, which is comparable to the previous estimation of the CME speeds obtained in \ref{Sec:WL_Observations} from \textit{STEREO A/COR 2} images.

%%%%%%%%%%%%%%%%%%%%%%%%%%%%%%%%%%%%%%%%%%%%%%%%%%%%%%%%%%%%%%%%%%%%%%
   \begin{figure}[h]
   \centering
   \begin{subfigure}[b]{0.49
   \textwidth}
    \includegraphics[width=\textwidth]{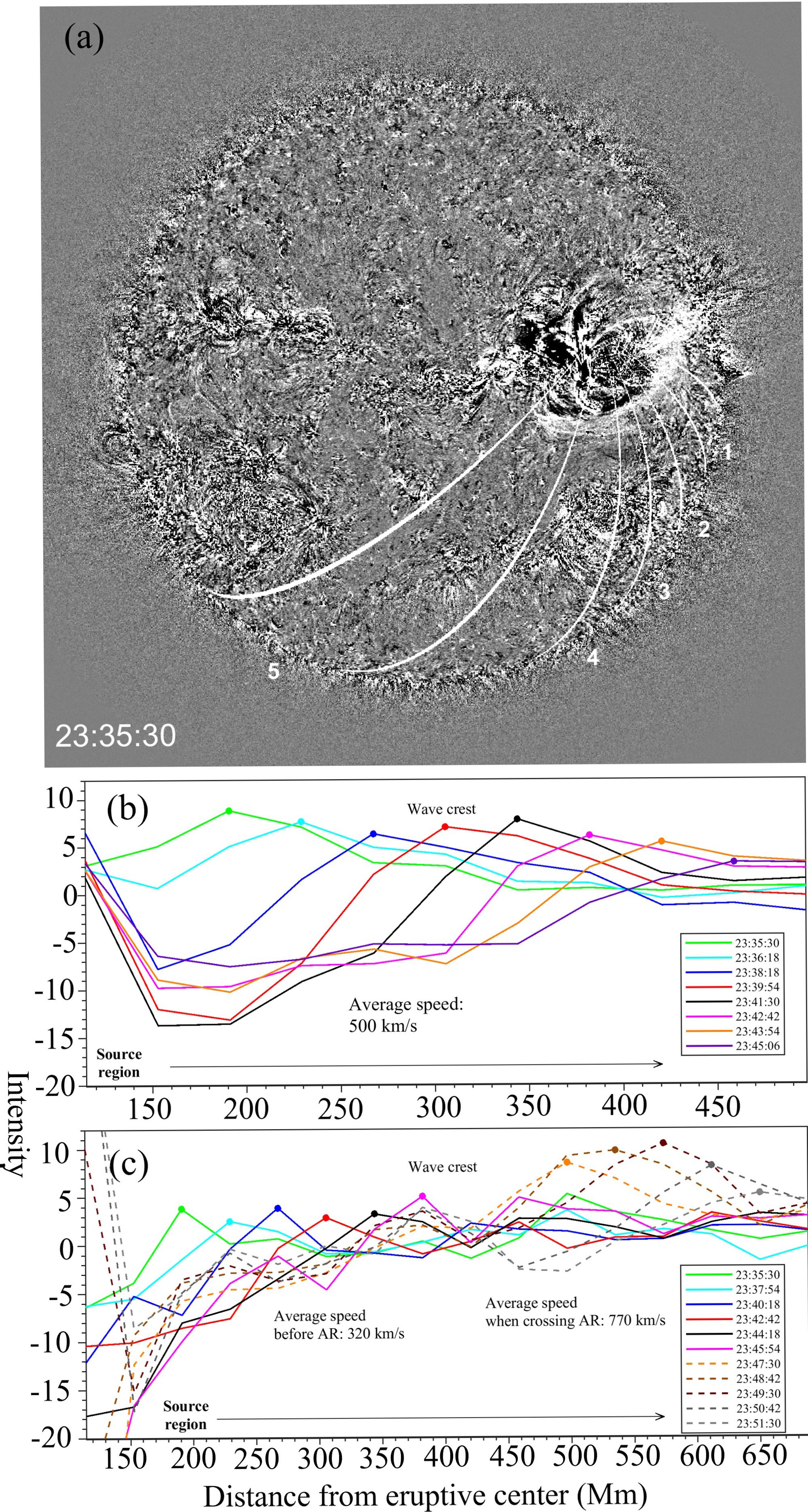}
   \end{subfigure}
%   \hfill
%   \begin{subfigure}[b]{0.49\textwidth}
%     \includegraphics[width=\textwidth]{Pictures/EIT_wave_profile.png}
%   \end{subfigure}
   \caption{ \textit{(a)} The EIT wave observed by the \textit{SDO/AIA} $193~\AA$ filter. The region where the EIT wave was most pronounced is divided into five sectors. \textit{(b)}Evolution of  the EIT wave profiles in sector 2, revealing a fast decay. \textit{(c)} Evolution of the EIT wave in sector 4.}
   \label{Fig:EUV_wave_analysis}
   \end{figure}
%%%%%%%%%%%%%%%%%%%%%%%%%%%%%%%%%%%%%%%%%%%%%%%%%%%%%%%%%%%%%%%%%%%%%%%%%%%%%%%%%

We use the EUropean Heliospheric FORecasting and Information Asset \citep[EUHFORIA;][]{Pomoell18} ideal-MHD heliospheric
model to study the CME propagation. The simulations were performed using the EUHFORIA v1.0.4 version of the model \citep{Hinterr19} The CME parameters such as the half width, direction of propagation (longitude and latitude) and 3D speed obtained from the GCS reconstruction were used as input for the cone CME model \citep[][]{Odstrcil96, Odstrcil99, Pomoell18, Scolini18}. The predictive capabilities of EUHFORIA were already described in \citep{Pomoell18, Scolini19, Scolini20}. 

Fig.~\ref{Fig:All_models}e shows the modelled CME after its insertion in the heliospheric domain. The ecliptic and the meridional cuts of the modelled CME and the background solar wind are shown in the left and right hand side of the figure, respectively. The CME then propagates self-consistently as a MHD disturbance (Fig.~\ref{Fig:All_models}d). The speeds of the modelled cone CME plotted in Fig.~\ref{Fig:All_models}e shows that the fastest component is close to the CME-flank regions.

\section{EIT wave} \label{Sec:EIT_Wave}

We studied the kinematics of the EIT wave associated with the CME using high cadence EUV imagery obtained by the \textit{SDO/AIA} 193~\AA filter. We derived the location and strength of the wave crest by calculating the intensity perturbation profiles from running difference image sequence using the ring analysis method \citep{Podladchikova2005,Podladchikova2019}.
We first constructed a spherical polar coordinate system with its centre on the brightest part of the associated flare, called the "eruptive centre" \citep[see e.g.][]{Warmuth04}. Then the image was divided into rings of equal width around the eruptive centre. We defined five angular sectors, where the EIT wave propagation is most pronounced (Fig. \ref{Fig:EUV_wave_analysis}a). Sectors 1 and 2 cover the regions of direct wave propagation, i.e. where it propagated without interactions, while sectors 3--5 are disturbed by strong interactions with ARs and the southern polar coronal hole. For each sector, we derived intensity perturbation profiles by calculating the mean intensity with the chosen binning of the rings. The outer border of every ring element is related with the corresponding distance from the source region. As a result, we obtained the projections of the radial intensity profiles onto the surface along the line-of-sight of \textit{SDO}.

Fig. \ref{Fig:EUV_wave_analysis}b and \ref{Fig:EUV_wave_analysis}c show the dependence of the EIT wave amplitude on the distance from the eruptive centre in sectors 2 and 4, respectively. Close to the source region, we observe areas of minimal intensity i.e., coronal dimming \citep[studied in][]{Veronig19}, which results from the density depletion caused by the evacuation of plasma during the CME lift-off \citep[e.g.,][]{Hudson1996,Thompson98,Dissauer2018}. The EIT wave front is characterised by a sharp increase of the intensity towards its maximum (wave crest) followed by a decay to the background level.
We identified the location of the wave crest (indicated by dots in Fig. \ref{Fig:EUV_wave_analysis}b, \ref{Fig:EUV_wave_analysis}c) over the period of EIT wave propagation from the eruptive centre towards the solar limb. The obtained mean velocity of the EIT wave in sectors 1 and 2, which are undisturbed by ARs, is 500 and 360 km/s, respectively. In sector 3, we observe a further decrease of the velocity to 310 km/s. However, when passing  through AR~11576 situated south from the source region, the EIT wave velocity doubled its value to about 720 km/s. This increase in the EIT wave speed is most probably related to the higher local Alfv\'en speed in regions of strong magnetic fields of the AR \citep[e.g.,][]{Mann99a}.

A similar profile is observed in sector 4 (Fig. \ref{Fig:EUV_wave_analysis}c), where the EIT wave propagated with a mean velocity of 320 km/s and accelerated to a speed of 770 km/s while passing through the southern AR. In sector 5, the wave is observed only as a quite diffuse structure. The average speed of the EIT wave was found to be about 280 km/s.
The results from the study of the EIT wave were used to reconstruct the coronal shock wave and study its propagation in Section \ref{Sec:EIT_association}.

%%%%%%%%%%%%%%%%%%%%%%%%%%%%%%%%%%%%%%%%%%%%%%%%%%%%%%%%%%%%%%
 \begin{figure}[h]
  \centering
  \begin{subfigure}[b]{0.49 \textwidth}
    \includegraphics[width=\textwidth]{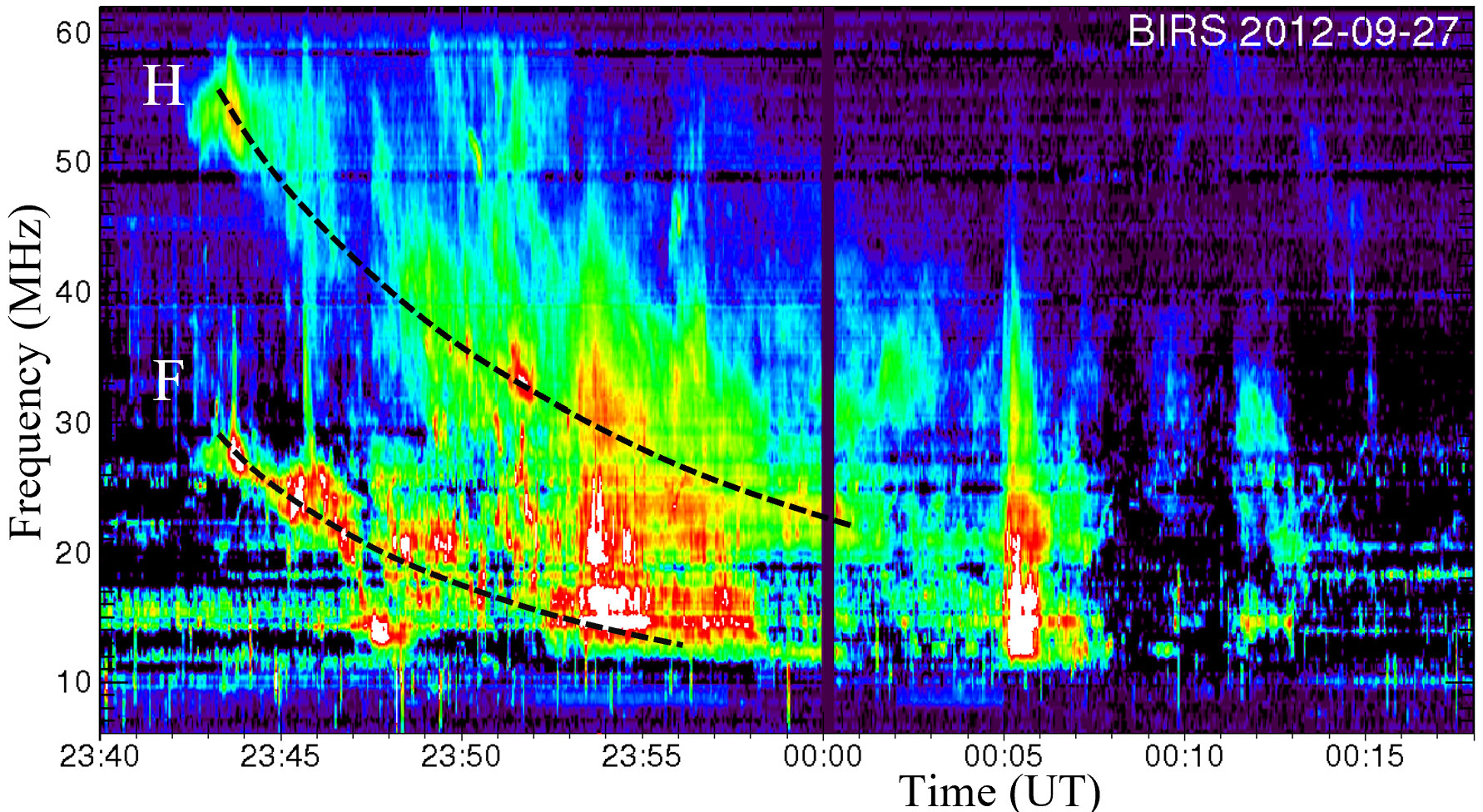}
  \end{subfigure}
  \caption{Dynamic spectra (observed by BIRS) shows radio emission in decametric and hectometric range. The dotted lines indicate the trend of the patchy type II burst observed at both fundamental and harmonic emission.}
  \label{BIRS}
  \end{figure}
%%%%%%%%%%%%%%%%%%%%%%%%%%%%%%%%%%%%%%%%%%%%%%%%%%%%%%%%%%%%%%%%

\section{Radio event} \label{Sec:Radio_analysis}

\begin{figure*}[h]
   \centering
   \begin{subfigure}[b]{0.85 \textwidth}
    \includegraphics[width=\textwidth]{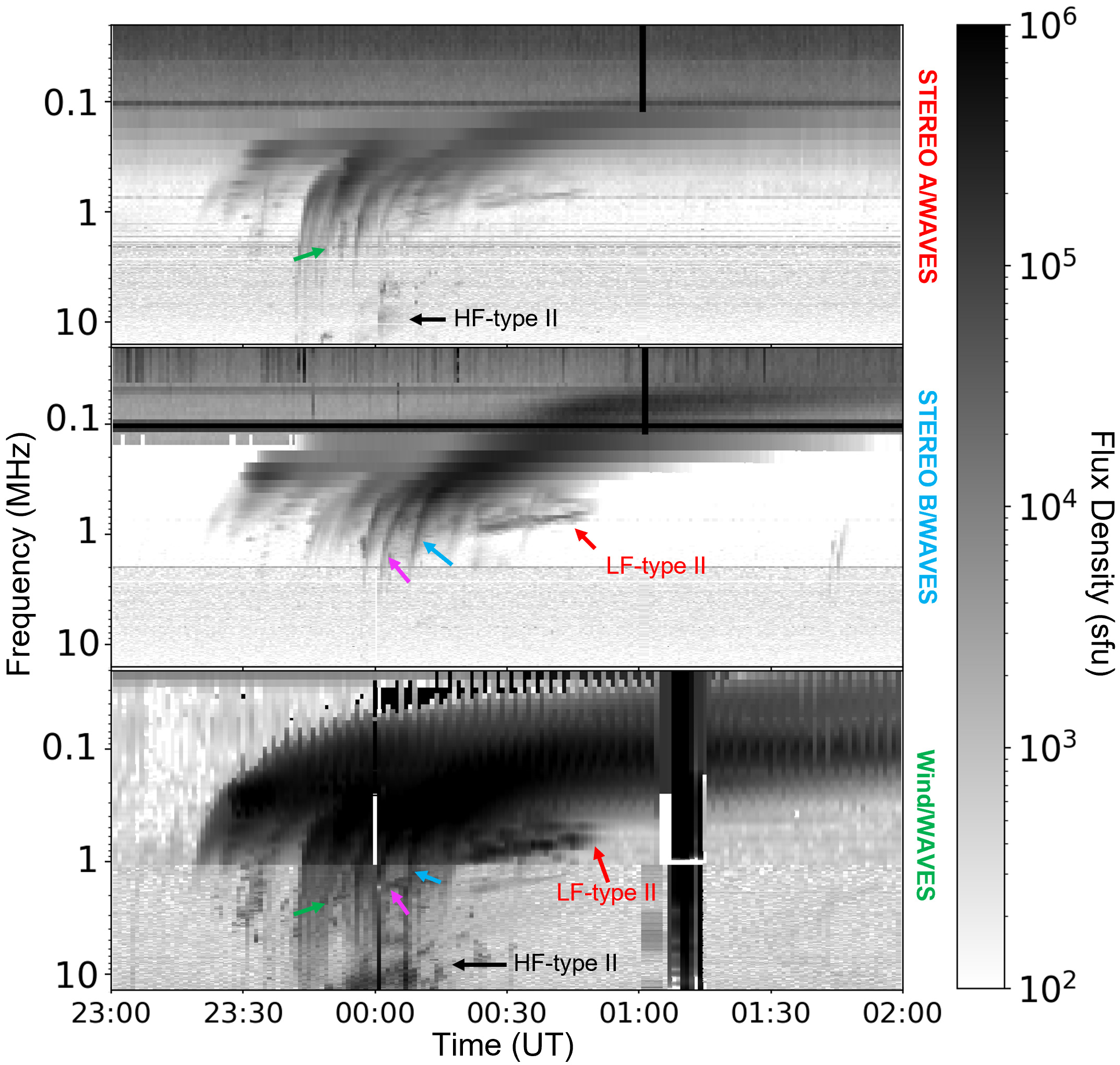}
   \end{subfigure}
   \caption{Calibrated dynamic radio spectra (solar flux units, i.e. \textit{sfu}), observed by the \textit{Wind/WAVES}, \textit{STEREO A/WAVES}, and \textit{STEREO B/WAVES} show the radio event associated with the flare/CME on September 27/28, 2012. The low frequency type II (LF-type II) burst indicated by the red arrow, observed by all three spacecraft, shows fundamental and second harmonic emission lane. The high frequency type II (HF-type II) burst which is indicated by the black arrow was best observed by \textit{Wind/WAVES}. The flare impulsive (FI) type III group (marked by the green arrow) and the type III associated with the flare decay (FD and FD*) phase are indicated by the blue and pink arrows, respectively.}
   \label{S_WAVES}
   \end{figure*}

The radio event associated with the studied CME/flare event was  observed by both, ground based and space based instruments. The BIRS dynamic spectra (Fig.~\ref{BIRS}) shows a structured and patchy metric type II burst which continues into the hectometric range as observed by the \textit{Wind/WAVES} and \textit{STEREO/WAVES} instruments (Fig.~\ref{S_WAVES}). This high frequency type II burst (hereafter, HF-type II) was observed from about 23:43~UT on September 27 until about 00:30 on September 28. A second type II was observed only by space based instruments due to its low starting frequency of 2000 ~kHz for \textit{Wind/WAVES}, and 1000~kHz as observed by the \textit{STEREO/WAVES} instruments. This low frequency type II (hereafter, LF-type II) was observed in the time interval 00:05\,--\,00:50~UT on September 28. Both the HF and LF-type II bursts show fundamental and second harmonic emission lanes. The LF-type II was observed by all three \textit{WAVES} instruments (on board \textit{Wind, STEREO A,} and \textit{STEREO B}). Groups of type III bursts (see Fig.~\ref{S_WAVES}) were also observed by all three WAVES instruments (during time interval 23:20\,--\,01:30), but with different starting frequencies as seen from different spacecraft. We note that the high frequency observations of \textit{STEREO B/WAVES} (up to 2~MHz) do not show any radio emission. Taking into account the spacecraft position at the time of the event (\textit{STEREO A} and \textit{STEREO B} separated by 125$^{\circ}$ and -118$^{\circ}$ from \textit{Wind}) and knowing that the source region of the CME/flare event was on the back side of the solar disk as observed by \textit{STEREO B}, we conclude that the radio emission was occulted for it. 

%%%%%%%%%%%%%%%%%%%%%%%%%%%%%%%%%%%%%%%%%%%%%%%%%%%%%%%%%%%%%%%%%%
\begin{figure}[h]
  \centering
  \begin{subfigure}[b]{0.49 \textwidth}
    \includegraphics[width=\textwidth]{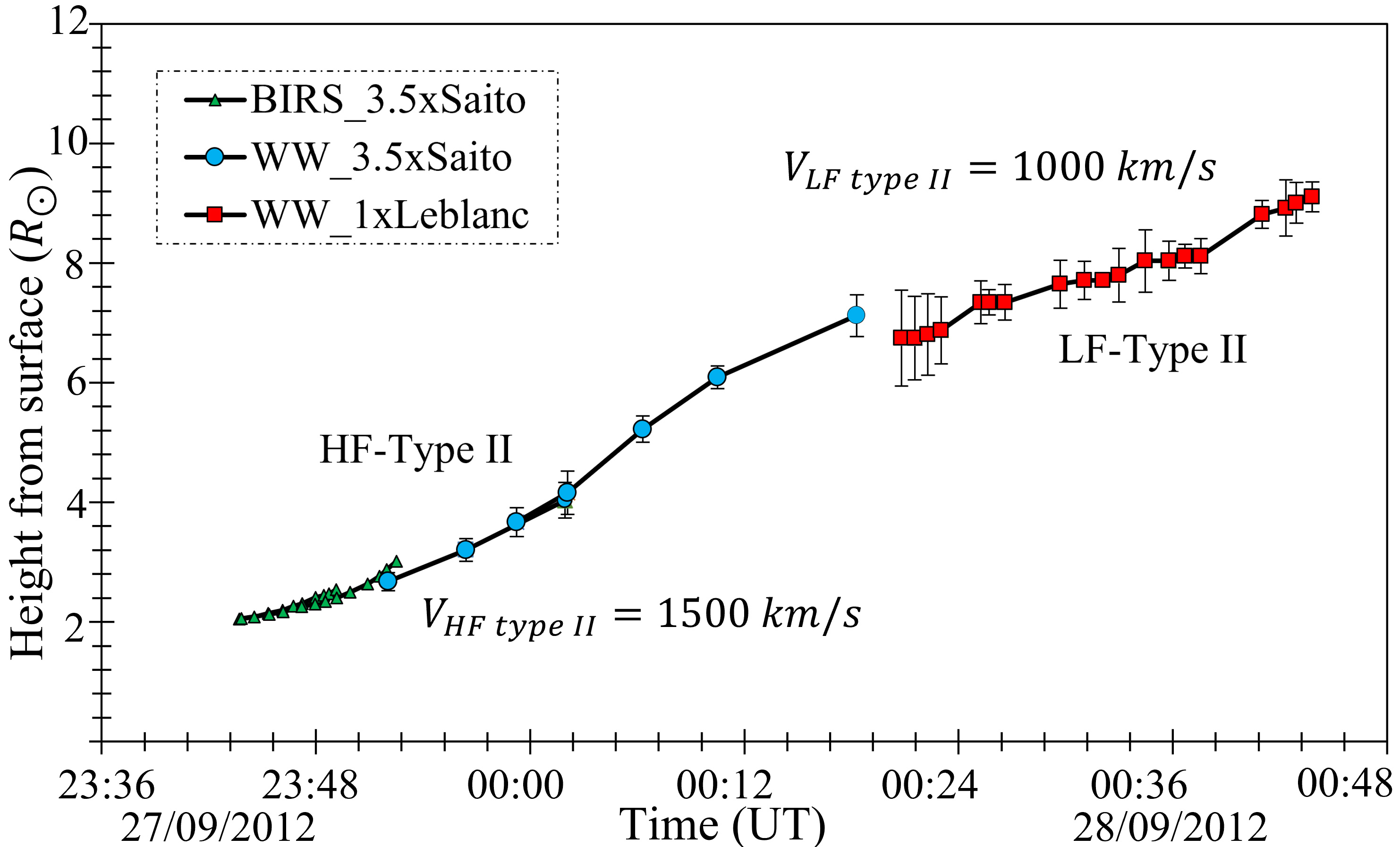}
  \end{subfigure}
  \caption{Kinematics of the shock wave obtained using the type II burst frequency drift and coronal electron density models. The high frequency type II (HF-type II) was observed by both ground and space based instruments, while the low frequency type II (LF-type II) was observed only by space based instruments. The HF-type II heights are obtained by using a 3.5-fold Saito density model, and the LF-type II heights are obtained using 1-fold Leblanc density model. The spectral range of the bursts are emphasised by the error bars }
  \label{Fig:Radio_kinematics}
  \end{figure}
%%%%%%%%%%%%%%%%%%%%%%%%%%%%%%%%%%%%%%%%%%%%%%%%%%%%%%%%%%%%%%%%%%%

%%%%%%%%%%%%%%%%%%%%%%%%%%%%%%%%%%%%%%%%%%%%%%%%%%%%%%%%%%%%%%%%%%%%%%%%%%%%%%%%%%%%%%
   
%%%%%%%%%%%%%%%%%%%%%%%%%%%%%%%%%%%%%%%%%%%%%%%%%%%%%%%%%%%%%%%%%%%%%%%%%%%%%%%%%%%%%%%
In order to obtain the type II kinematics, we employed the classical method \citep[e.g.,][]{Magdalenic08, Magdalenic10, Magdalenic14} using the drift rate of the radio bursts and coronal electron density models. The \cite{Saito70} and \cite{Leblanc98} electron density models are two of the most frequently employed 1D density models for the metric and decametre to hectometre frequency range, respectively. Similar to some previous studies we employ a 3.5-fold Saito density model for metric observations \citep[e.g.][]{Magdalenic08,Magdalenic10} and a 1-fold Leblanc density model for for the decametre to hectometre range \citep[e.g.][]{Palmerio19}. Fig.~\ref{Fig:Radio_kinematics} shows the type II drift rates estimated by considering the central part of the emission band. Type II speeds obtained with this method for the HF-type II and the LF-type II are about 1500~km/s and 1000~km/s, respectively. The error bars in Fig.~\ref{Fig:Radio_kinematics} show the uncertainty of the obtained results. Fig.~\ref{Fig:Radio_kinematics} indicates that the LF-type II burst is the continuation of the HF-type II burst. However, their strongly different positions in the dynamic spectra (Fig.~\ref{S_WAVES}) do not support this conclusion.
In Fig.~\ref{Fig:CME_Kinematics}a, we present the LF-type II kinematics employing three different 1D density models in order to highlight the drastic change in interpretation of the radio emission induced by different models.

A more accurate method to estimate the shock wave kinematics is the so-called radio triangulation that employs direction-finding observations (Sec. \ref{Sec:Radio_triangulation}).
However, as the direction-finding observations are not always available, we will first discuss the radio event qualitatively. A more quantitative way using radio triangulation is presented in Sec. \ref{Sec:typeII_triangulation}.
The \textit{Wind/WAVES} observations (Fig.~\ref{S_WAVES}, bottom panel) show both fundamental and the second harmonic lanes of intense LF-type II burst. A qualitative assessment indicates that the intensity of the radio burst is strongest in \textit{Wind/WAVES}, somewhat fainter in the \textit{STEREO B/WAVES}, and very faint in the \textit{STEREO A/WAVES} observations (Fig.~\ref{S_WAVES} middle and top panel, respectively). Taking into account the assumption that the radio emission is most intense in the direction of its propagation \citep[like e.g. in][]{Magdalenic14}, we can roughly deduce the direction of the shock wave propagation to be between \textit{STEREO B} and \textit{Wind}, and somewhat closer to the \textit{Wind} spacecraft. Although this is only a qualitative assessment, it can provide additional information in a case when only the classical method for estimation of the shock wave kinematics is possible.

% The kinematics of the type II radio emissions from all observations are plotted in figure (4) 

%%%%%%%%%%%%%%%%%%%%%%%%%%%%%%%%%%%%%%%%%%%%%%%%%%%%%%%%%%%%%
  \begin{figure*}[!h]
   \centering
    \includegraphics[width=0.89\textwidth]{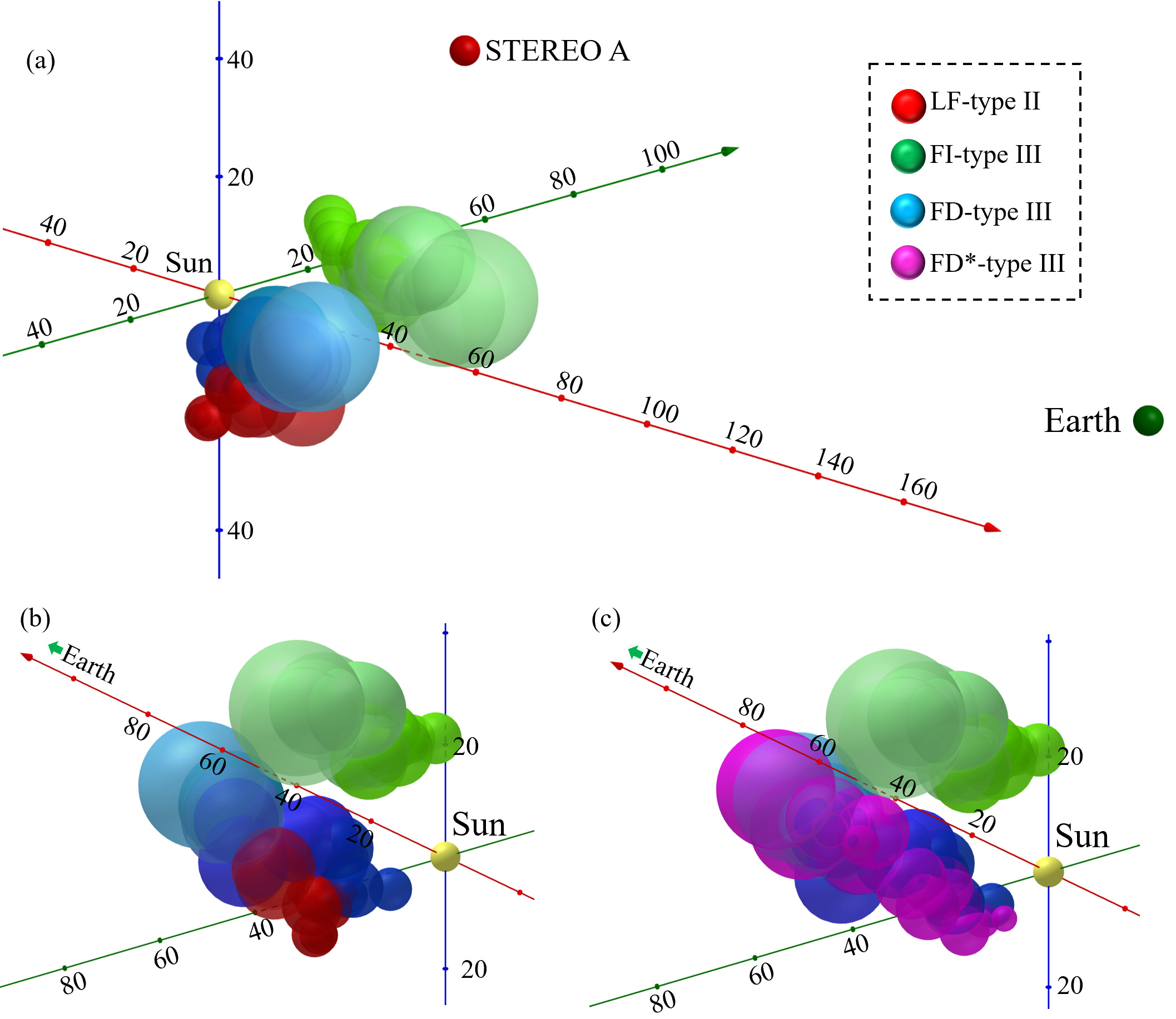}
    \caption{(a) Results from radio triangulation studies of type II and type III radio bursts. The yellow sphere represents the Sun, while the red and green spheres represent \textit{STEREO A, STEREO B} and \textit{Wind}, respectively. The spheres of varying size represent the radio source regions. The green spheres mark the radio source positions of the flare impulsive type III (FI-type III) close to the CME flank. The low frequency type II (LF-type II) and the flare decay type III (FD-type III), denoted by the red and blue spheres, have source positions at the south-east side of the Sun, i.e. close to another CME flank. (b) The FI-type III, FD-type III and LF-type II plotted from a different perspective. (c) Three different type IIIs plotted together.}
   \label{Fig:all_triangulation}
   \end{figure*}
%%%%%%%%%%%%%%%%%%%%%%%%%%%%%%%%%%%%%%%%%%%%%%%%%%%%%%%%%%%%%

\section{Radio Triangulation} \label{Sec:Radio_triangulation}

The kinematics obtained from type II drift rate using radial density models, as the one presented in Fig.~\ref{Fig:CME_Kinematics}a, is useful but does not provide information on the spatial position of the radio sources. Therefore, we will use the unique method for estimating the 3D positions of the radio sources in the interplanetary space, the so called radio triangulation technique. This technique was so far mostly used to study type III radio bursts \citep[][]{Fainberg72, Gurnett78, Reiner88}, and only recently it has been more frequently used to study type II bursts \citep[][]{Hoang98, Reiner98, MartinezOliveros12, MartinezOliveros15, Magdalenic14, Krupar16, Makela16, Makela18, Krupar19}. Depending on the type of spacecraft, spinning or three axis stabilized, we distinguish different direction finding techniques \citep[e.g.,][]{Fainberg74, Lecacheux78, Manning80, Santolik03, Cecconi05, Krupar12, MartinezO12}. For \textit{Wind} (spinning spacecraft) observations we employed a spin demodulation technique \citep[][]{Fainberg74} and for \textit{STEREO} (three axis stabilised spacecraft) observations we employed a singular value decomposition technique \citep[][]{Krupar12}. The radio triangulation studies are performed using simultaneous direction-finding observations of at least two spacecraft. 

The radio triangulation analysis in this study was done employing the following premises:

\begin{itemize}
    \item[\textbullet]
    The direction finding observations are available for a selected set of frequency channels at each spacecraft. The observing frequencies of \textit{STEREO} and \textit{Wind} are slightly different \citep{Bougeret95, Bougeret08} which might induce uncertainty in the radio triangulation results. Similar to previous studies \citep[e.g.][]{MartinezOliveros12, Magdalenic14, MartinezOliveros15, Makela16, Makela18}, for combining \textit{Wind} and \textit{STEREO} observations we considered the closest frequency pairs.

    \item[\textbullet]
    The direction finding technique provides wave vectors which are used in the radio triangulation studies to estimate the 3D radio source positions. As previously done in \cite{Magdalenic14}, we use the full distance between the two wave vectors, at a given frequency pair, as the radio source region. This region is then presented in the figures as a sphere, with the diameter equal to the the distance between the wave vectors. We note that the intrinsic geometric errors of the radio triangulation technique are quite large regardless of frequency, and they are mostly due to the receiver gain and the position of the spacecraft pair \citep[see][for more details]{Krupar12}. Therefore, we do not discuss the geometric radio source sizes, but only the radio source regions as defined above. A Similar procedure was used in previous studies \citep{Reiner98, MartinezOliveros12, MartinezOliveros15, Makela16, Makela18, Krupar20}

    \item[\textbullet]
    The estimated distance between the two wave vectors is generally smaller at high frequencies. Hence, In this analysis we used the highest available frequencies of direction finding observations, and we did not use frequencies below 500 kHz.
    
    \item[\textbullet]
    The points for radio triangulation studies were selected taking into account the time delay which is due to different travel-times needed for a radio signal to arrive at the different spacecraft. The magnitude of the time delay is not absolute as it depends on the direction of propagation of the radio emission.

    \item[\textbullet]
    The intensity of the type II bursts is significantly lower than for the type III bursts, so  in order to have the radio emission sufficiently above the background level, similar to \cite{Magdalenic14}, we employed background subtraction of only 5\% for all direction-finding data. 

\end{itemize}

Three combinations of direction finding observations are possible. The results obtained using \textit{STEREO A/WAVES} and \textit{STEREO B/WAVES} observations are unreliable due to large angular separation of the spacecraft. We present the results for the other two spacecraft pairs; \textit{STEREO A/WAVES} and \textit{Wind/WAVES}, and the \textit{STEREO B/WAVES} and \textit{Wind/WAVES}. These direction-finding observations not only show the highest intensity of radio flux but also give the smallest distances between the wave vectors and the most reliable results. 

\subsection{Source positions of type III radio bursts} \label{Sec:typeIII_triangulation}

We distinguish two groups of type III bursts associated with the studied event. The first group are type III bursts temporally associated with the flare impulsive phase (FI), observed at about 23:52 UT (marked in Fig. \ref{S_WAVES} by green arrow). The second group are the two type III bursts associated with the flare decay phase (FD), observed at 00:00 and 00:10 UT (marked in Fig. \ref{S_WAVES} by blue and pink arrows). Similar to the qualitative analysis of the type II bursts, (see, Sec. \ref{Sec:Radio_analysis}), we also discuss propagation of the type III bursts. If we consider that the radio emission is the most intense in the direction of its source propagation, and use only the dynamic spectra, we can deduce that the source region of FI-type III appears closer to \textit{STEREO A} than the source region of the FD-type IIIs (FD-type III and FD*-type III). Likewise, the source region of the FI-type III appears to be further away from \textit{STEREO B} than the source region of the FD-type III bursts. In order to quantify the possibly different source positions of the FI-, FD-type III, and FD*-type III bursts, we performed radio triangulation.

For triangulating type III bursts we used observations from two spacecraft pairs: a) \textit{Wind/WAVES} and \textit{STEREO A/WAVES}; and b) \textit{Wind/WAVES} and \textit{STEREO B/WAVES}. For all type III bursts, we considered the same frequency pairs (in kHz): 525/548, 575/548, 625/624, 675/624, 725/708, 775/708, 825/804, 925/916, 1025/1040 and 1075/1040, respectively.

%%%%%%%%%%%%%%%%%%%%%%%%%%%%%%%%%%%%%%%%%%%%%%%%%%%%%%%%%%%%%  

  \begin{figure}[!h]
   \centering
   \begin{subfigure}[b]{0.49\textwidth}
    \includegraphics[width=\textwidth]{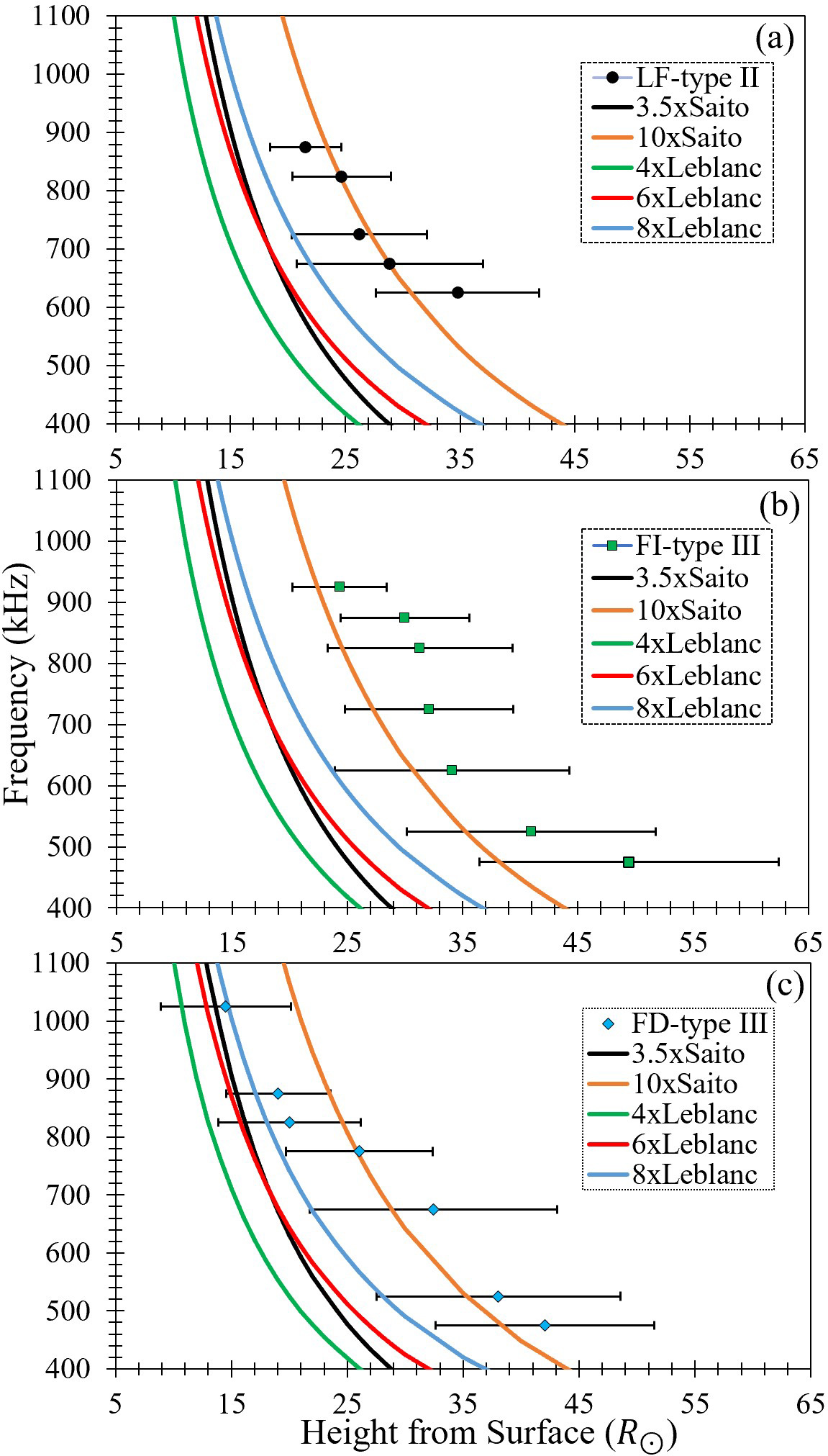}
   \end{subfigure}
   \caption{Frequency shown as a function of heliocentric distance. Radio triangulation results are plotted together with two density models (Saito, Leblanc) for comparison. The markers show the source positions of three different radio bursts and the bars attached to them show the distance between the wave vectors. (a) The black spheres show LF-type II, (b) FI-type III is noted by the green squares, and in (c) the blue diamonds indicate the FD-type III.}
   \label{Fig:triang_density}
   \end{figure} 
%%%%%%%%%%%%%%%%%%%%%%%%%%%%%%%%%%%%%%%%%%%%%%%%%%%%%%%%%%%%%

The results of the triangulation are shown in Fig.~\ref{Fig:all_triangulation}a. The yellow sphere represents the Sun, and red, blue and green ones represent the three spacecraft \textit{STEREO A, STEREO B} and \textit{Wind}, respectively. The source positions of radio bursts are colour coded. Darker colours denote sources situated closer to the Sun (high frequencies) and the lighter coloured ones are further away from the Sun (low frequencies). The type II source positions are denoted as red spheres and the type III source positions are denoted as green, blue, and pink spheres.

Fig.~\ref{Fig:all_triangulation}b shows that the sources positions as well as propagation path of the FD-type III
and FD*-type III are significantly different from the FI-type III. The open field lines, along which the FD-type III and FD*-type III bursts propagate, are in the south-west quadrant of the Sun. The FI-type III bursts were observed in the north-west quadrant of the Sun. The change in the type III source positions happens at about the flare peak time. We note that the direction-finding observations allowed us, for the first time, to quantitatively estimate the significantly different source positions of type III bursts associated with one CME/flare event.

\subsection{The low frequency type II radio burst} \label{Sec:typeII_triangulation}

The radio triangulation study was only performed for the LF-type II burst, because the HF-type II was not observed in the range of the direction-finding frequencies (Fig.~\ref{S_WAVES}). For the analysis we selected the following frequency pairs (in kHz): 575/548, 625/624, 675/624, 725/708, 775/708, 825/804, 875/804 and 925/916 from \textit{STEREO B} and \textit{Wind}, respectively.

%%%%%%%%%%%%%%%%%%%%%%%%%%%%%%%%%%%%%%%%%%%%%%%%%%%%%%%%%%%%%%%%%%%%%%%%%%%%%%%%%%%%%%%%%%%%

\begin{figure*}[h!]
     \centering
     \begin{subfigure}[b]{0.9\textwidth}
         \centering
         \includegraphics[width=\textwidth]{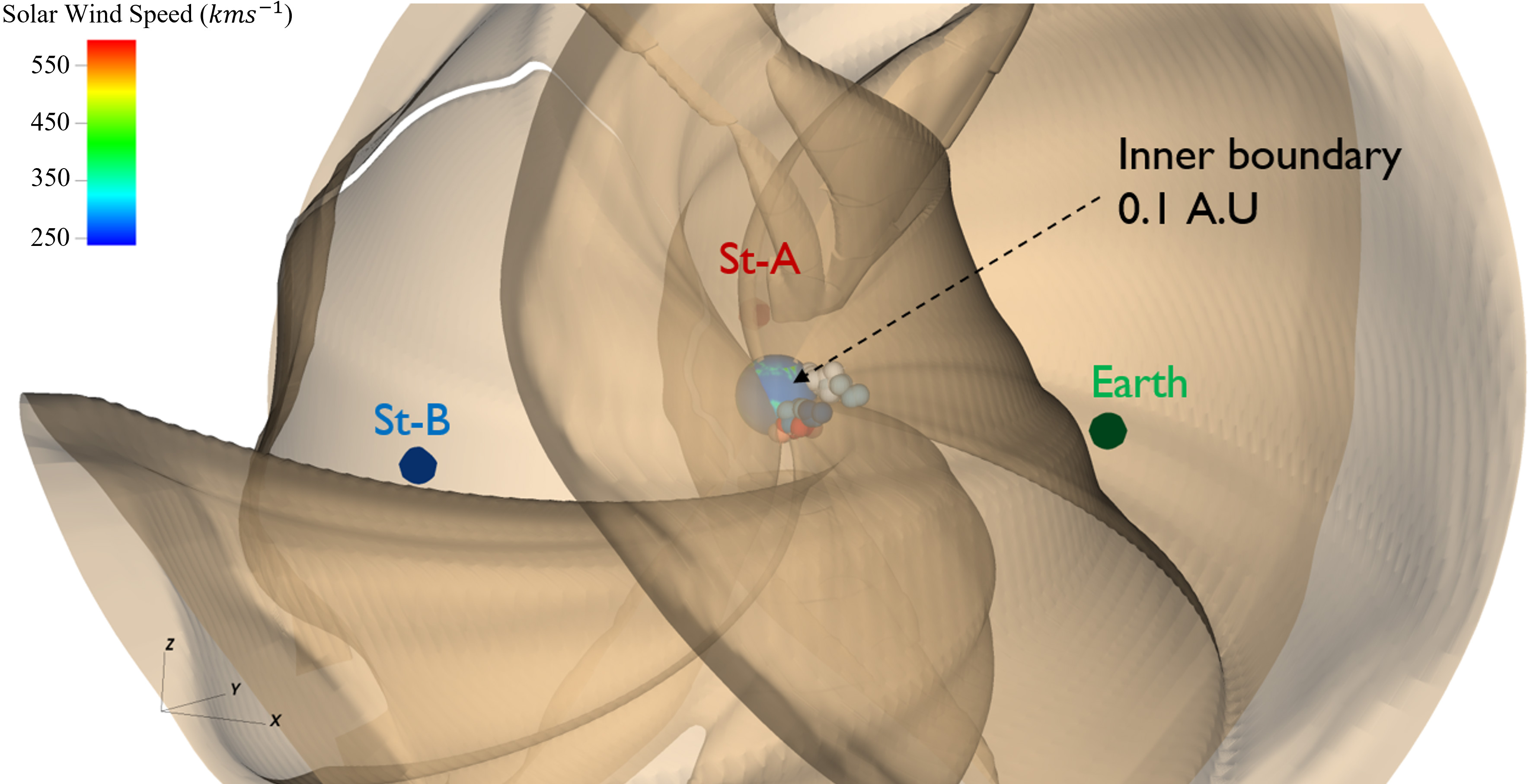}
     \end{subfigure}
     \caption{Radio triangulation results plotted together with the isosurface of the heliospheric current sheet ($B_{r} = 0$) modelled by EUHFORIA. The radio sources of the LF-type II emission seem to cross the complex structure of the 
     heliospheric current sheet several times.}
     \label{Fig:HCS}
     
\end{figure*}

%%%%%%%%%%%%%%%%%%%%%%%%%%%%%%%%%%%%%%%%%%%%%%%%%%%%%%%%%%%%%%%%%%%%%%%%%%%%%%%%%%%%%%%%%%%%%%%%

Fig.~\ref{Fig:all_triangulation}a shows that the source positions of the LF-type II are situated in the south-west part of the Sun. The darker coloured circles represent high frequency pairs positioned closer to the Sun. Close up shows the slow drift of the type II emission from the south towards the solar central meridian (Fig.~\ref{Fig:all_triangulation}b). We also note that the LF-type II positions are roughly co-spatial with the positions of the FD-type III burst. 

\subsection{Coronal electron density profiles and propagation direction of the radio emission} \label{Sec:density_triangulation}

The 3D source positions of the radio emission obtained from the triangulation study can be converted to radial distances (Fig.~\ref{Fig:triang_density}), and compared with the generally employed 1D coronal density models. The radio source positions are plotted in Fig.~\ref{Fig:triang_density} together with 1D coronal electron density profiles \citep{Saito70, Leblanc98}. The frequency ($f \propto n$) is presented as a function of the radial heights. The horizontal bars denote distances between the two wave vectors. 

The obtained density profiles along the propagation path of the LF-type II (Fig. \ref{Fig:triang_density}a) and FD-type III (Fig. \ref{Fig:triang_density}c) are similar, crossing different density models (from 3.5-fold Saito to 8-fold Leblanc). The similarity of the profiles is expected as the source regions of these bursts are propagating through the same region in the corona. The trend of crossing different density models is probably a consequence of the non-radial propagation of the radio source (Fig.~\ref{Fig:Projection_tpyeII}). The FI-type III (Fig. \ref{Fig:triang_density}b) shows a somewhat different profile, which is expected as the electron beam propagated along a magnetic field line on a different flank of the CME (Fig. \ref{Fig:all_triangulation}). We note that all density profiles obtained from the radio triangulation study indicate unusually high values. We note that a 10-fold Saito density model is used rarely and only in cases of large eruptions in the low corona (\(\approx 1.5~R_{\odot}\)), like e.g. \cite{Pohjolainen08b}. One of the processes which possibly influences the results of the radio triangulation is the scattering of the radio emission \citep[see e.g.][]{Melrose70, Thejappa07, Kontar19}. We do not exclude that the absolute values of herein obtained densities might be impacted by scattering effects.
Although the scattering can indeed influence the observed radio source positions, it should not significantly affect the general direction of the propagation of the radio sources. We believe that the obtained results are mainly due to the non-radial propagation directions of the radio sources.

\section{Ambient coronal conditions and their influence on the eruptive event} \label{Sec:Coronal_conditions}

\subsection{Shock wave propagation through the corona - interaction with streamer} \label{Sec:streamer_interaction}

\begin{figure*}[h!]
     \centering
     \begin{subfigure}[b]{0.99\textwidth}
         \centering
         \includegraphics[width=\textwidth]{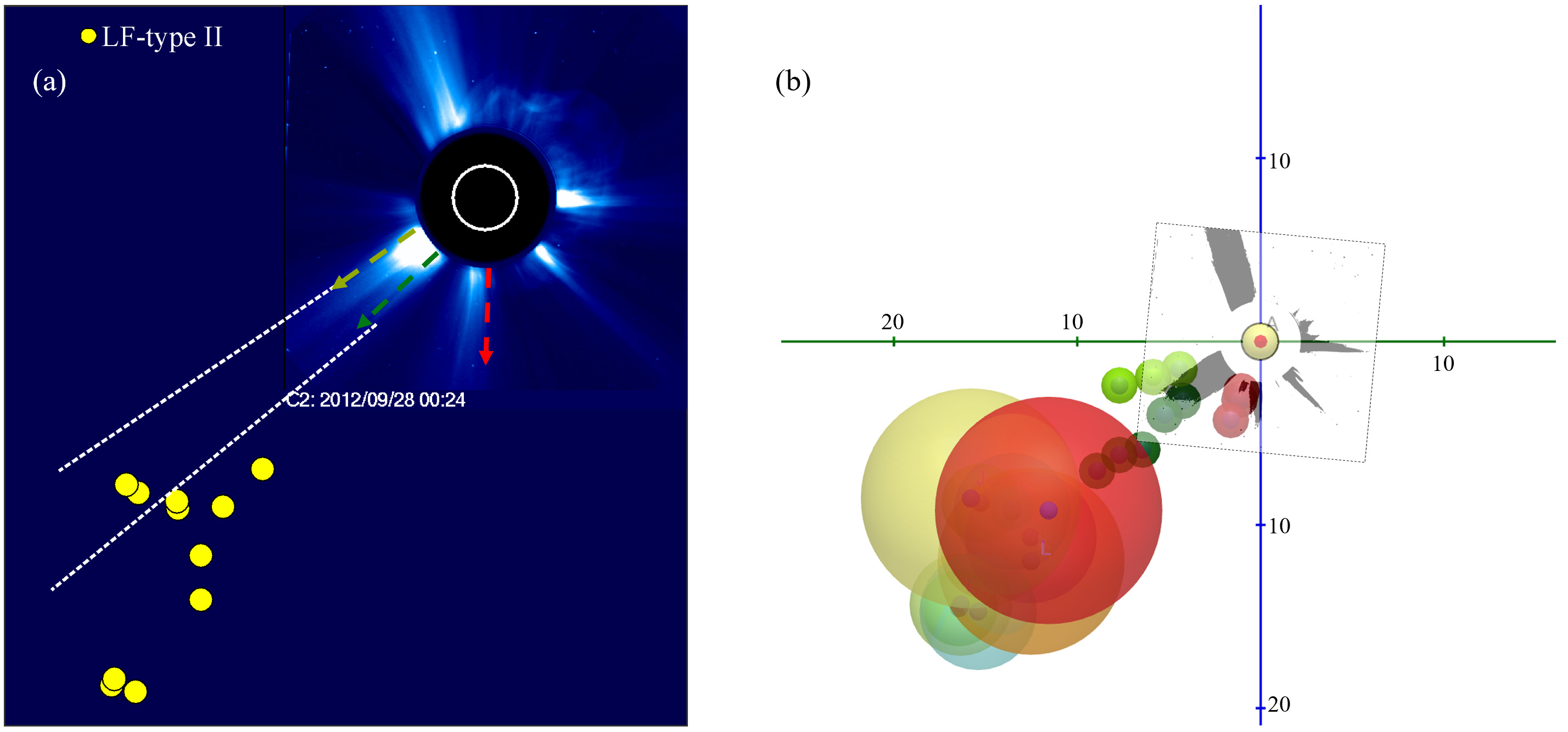}
     \end{subfigure}
     \caption{\textit{(a)} Centroids of the LF-type II emissions projected on a \textit{SOHO/LASCO C2} image shows their proximity to the streamer stalk region in the plane of sky, giving an indication on the probable shock/streamer interaction. Colored arrows are the streamer regions selected for reconstruction. \textit{(b)} Reconstructed streamers (color coded) plotted together with the type II source regions (distance between the wave vectors) in a plane of sky projection similar to the \textit{SOHO/LASCO C2} image. The different frequency pairs of the type II source regions are denoted by a rainbow color scale (Blue hues: high frequency pairs, Red heues: low frequency pairs). The source regions of the type II burst are presented as spheres. The smaller spheres (green, olive, and red) are the positions of the reconstructed streamers.}
     \label{Fig:Streamer_LASCO_projection}
\end{figure*}

We also investigated the possibility of the shock wave interactions with the ambient coronal structures. \ref{Fig:HCS} shows how EUHFORIA models the heliospheric current sheet in the time of the studied event. The complex structure of the heliospheric current sheet (HCS) is not unusual during high levels of solar activity. Employing the coronagraph observations we identified three streamers in the south-east quadrant of the Sun (Fig.~\ref{Fig:Streamer_LASCO_projection}a) that were perturbed by the passage of the shock wave. The bending of the streamers due to the shock wave propagation was particularly well visible in white-light coronagraph observations by \textit{STEREO B/COR 1}).

We have found that the direction along which the LF-type II emission was located, coincided with the direction of the fastest EIT wave component (Section \ref{Sec:EIT_Wave}, Fig.~\ref{Fig:EUV_wave_analysis}), i.e. the south-east from the source region. In order to understand the relative position of the LF-type II burst and the nearby coronal structures, in particular streamers \citep[a preferable place for the generation of radio emission,][]{Shen13, Floyd14} we projected the centre of the radio source region on the \textit{SOHO/LASCO C2} image. A white-light image recorded at 00:24~UT, shortly after the CME eruption, was selected. Most of the projected LF-type II sources were outside the range of the \textit{SOHO/LASCO C2} field of view (Fig.~\ref{Fig:Streamer_LASCO_projection}a), therefore for better comparison we mark the edges of the streamer stalk region at the heights beyond the \textit{SOHO/LASCO C2} field of view. Fig.~\ref{Fig:Streamer_LASCO_projection}a shows that the projected positions of the centres of the radio source regions and the streamer stalk are close to each other.

%%%%%%%%%%%%%%%%%%%%%%%%%%%%%%%%%%%%%%%%%%%%%%%%%%%%%%%%%%%%%%%%%%%%%%%

%%%%%%%%%%%%%%%%%%%%%%%%%%%%%%%%%%%%%%%%%%%%%%%%%%%%%%%%%%%%%%%%%%%%%%%%

We also performed a 3D reconstruction of the streamers at the south-east quadrant of the Sun using the tie-pointing method \citep[e.g.][]{Inhester06}. Fig.~\ref{Fig:Streamer_LASCO_projection}b shows that the 3D positions of the reconstructed streamer are in agreement with the projected LF-type II source regions. We note that this way of projecting the type II sources, from 3D space to the 2D plane of sky, made the sources to be apparently stationary. This effect is less visible in the Fig.~\ref{Fig:Streamer_LASCO_projection}a in which only the position of the source region centres, and not the full distance between the wave vectors, is presented. All this, together with the knowledge of the 3D position of radio sources allows us to suggest, similar to previous studies \citep{FengSW12, Magdalenic14, Zucca18, Mancuso19}, that the type II radio emission was enhanced by the interaction between the shock wave and the streamer.

%\textbf{ We note that the straightforward association %of the LF-type II positions and the EIT wave can be %made only employing MHD modelling of the EIT wave and %shock wave.}

\subsection{Shock wave propagation through the corona - association with the EIT wave} \label{Sec:EIT_association}

The EIT wave study (Section~\ref{Sec:EIT_Wave}) shows that the speed of the EIT wave is larger when considering the directions from the source region towards the southern polar coronal hole (direction 2 and 3 in Fig.~\ref{Fig:EUV_wave_analysis}). If we assume that the EIT wave is the low coronal counterpart of the coronal shock wave \citep[e.g.][]{Mann99b, Warmuth05, Veronig06, Muhr10, Warmuth15}, then the propagation direction of the fastest component of the EIT wave should roughly correspond to that of the fastest component of the CME and associated shock wave (as shown in Section \ref{Sec:CME_modelling}, Fig.~\ref{Fig:All_models}e). This region also coincides with the one where the LF-type II sources are situated.

In order to further inspect the association of the EIT wave and type II bursts we performed a simple 3D-reconstruction of the EIT wave \citep[similar to,][]{Zucca14, Zucca18, Rouillard16} using parameters obtained in Section \ref{Sec:EIT_Wave}, and the global magnetic field configuration using a Potential Field Source Surface model \citep[\textit{PFSS};][]{Schrijver03}. In the presented model of the EIT wave in the 3D domain, we restricted to the heights of \( 2.5 R_{\odot} \) in order to avoid oversimplification of the wave dynamics at the larger heights as the anisotropic wave expansion might result in a wave deformation \citep{Temmer11}. Due to this height restrictions, the modelled results are constrained to the low corona and can directly be compared only with the HF-type II burst. Nevertheless, the model can give an indication on the possible shock region associated with the LF-type II burst.
%We consider that the EIT wave is the low coronal counterpart of the shock wave observed higher in the corona.

%The height corresponding to the starting frequency of the HF-type II is \( 1.8 R_{\odot} \) and was based on the classical analysis using a 3.5-fold Saito coronal electron density model (Fig.~\ref{kinematics}). 

\begin{figure*}[ht]
     \centering
     \begin{subfigure}[b]{0.99\textwidth}
         \centering
         \includegraphics[width=\textwidth]{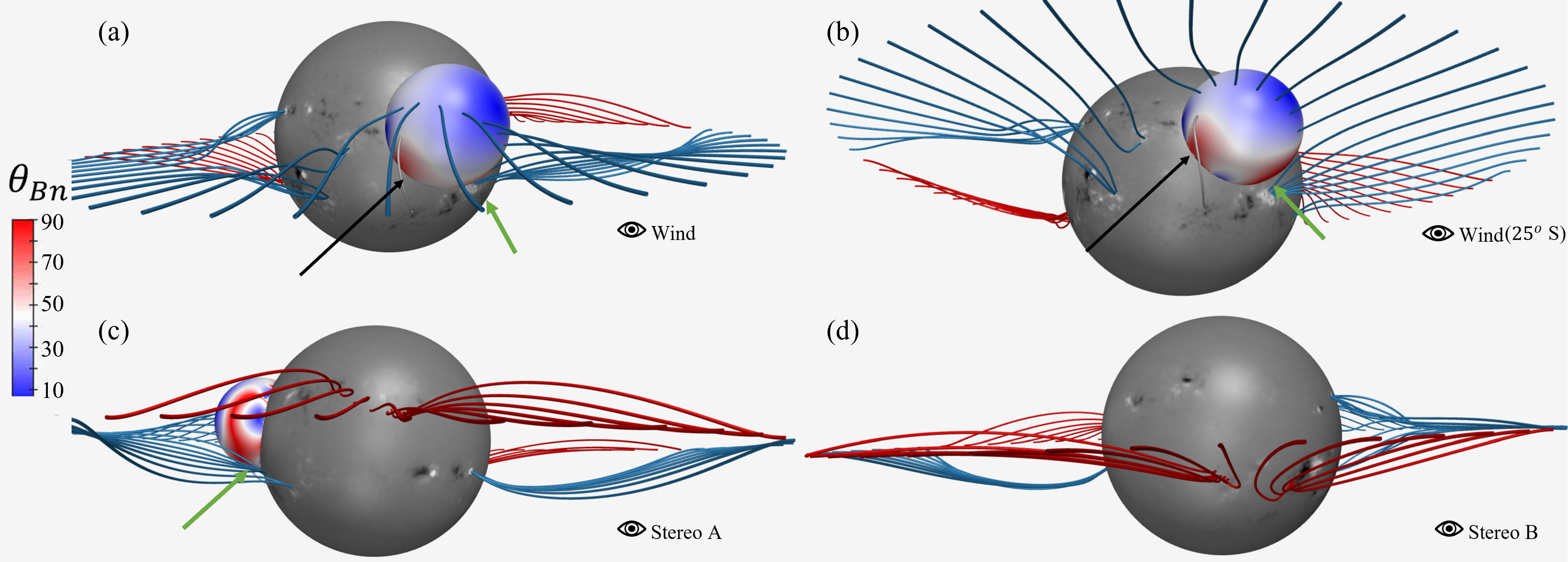}
     \end{subfigure}
     \hfill
     \caption{The reconstructed EIT wave bubble and its corresponding high coronal expansion using the measurements made in the analysis described in Section \ref{Sec:EIT_Wave}. The panels show the event as observed from the point-of-view of different spacecraft. \textit{(a)} shows the EIT wave observed by \textit{SDO/AIA}, and \textit{(b)} when viewed from $25^{o}$ south. \textit{(c)} shows the \textit{STEREO A} point of view, and \textit{(d)} by \textit{STEREO B}. The coronal magnetic field was extrapolated using a PFSS model. The field line colours correspond to their polarity, with red being positive and blue negative. The shock wave bubble is plotted at the start time of the HF-type II (23:44~UT) and the colours on the surface of the bubble are the values of the shock normal angle (\( \theta_{Bn} \)) to the local magnetic field. The green arrow indicates the possible source region of the HF-type II while the black arrow is the possible region of the part of the shock wave which corresponds to the LF-type II burst.}
     \label{Fig:shock_bubble}
\end{figure*}

Fig.~\ref{Fig:shock_bubble} shows the reconstructed dome of the EIT wave at the start time of the HF-type II burst (23:45~UT on September 27). The EIT wave dome shows the quasi-perpendicular shock normal angle (\( \theta_{Bn} \)) in the south-east and south-west region (marked in Fig.~\ref{Fig:shock_bubble} by black and green arrows, respectively). We believe that the south-western region (green arrow in Fig.~\ref{Fig:shock_bubble}) is the most probable source location of the HF-type II burst. This conclusion also agrees with the so called intensity-directivity relationship of the radio emission \citep[][]{Magdalenic14}. In brief, the intensity of the HF-type II is strongest as seen by \textit{Wind/WAVES}, weak as seen by \textit{STEREO A/WAVES} and not observed by \textit{STEREO B/WAVES}. This suggests that the source of the HF-type II is fully occulted for the \textit{STEREO B/WAVES} and it propagates mostly in the direction of \textit{Wind/WAVES}. 

%%%%%%%%%%%%%%%%%%%%%%%%%%%%%%%%%%%%%%%%%%%%%%%%%%%%%%%%%%%%%%%%%%

%%%%%%%%%%%%%%%%%%%%%%%%%%%%%%%%%%%%%%%%%%%%%%%%%%%%%%%%%%%%%%%%%%

The south-east region, marked by black arrow in Fig.~\ref{Fig:shock_bubble}, agrees with the positions of the LF-type II source regions obtained by radio triangulation (at larger heights) and with the fastest CME segments, as modelled by EUHFORIA at about \( 30 R_{\odot} \) (Fig. \ref{Fig:All_models}d). Taking all this into account, the assumption that the conditions for the quasi-perpendicular regime in the south-east region (as modelled in the low corona, Fig. \ref{Fig:shock_bubble}), are also met at larger heights is reasonable. However, if this is not the case, the interaction of the shock wave and streamer
can provide an additional favourable condition for the generation of the shock associated radio emission \citep[as already shown by][]{Shen13, Magdalenic14, Zucca18}. Previous studies \citep[e.g.][and references therein]{Holman83, Mann95book, Reiner98, Mann03, Mann05} have demonstrated that a quasi-perpendicular shock wave geometry is significantly more efficient in accelerating particles and therefore producing radio emission. Some studies have indicated that a shock wave can be radio quiet in the sub-critical regime and produce radio emission when super-critical \citep{Gopalswamy10, Gopalswamy12}. Even so, the favourable conditions are not only the quasi-perpendicular shock wave geometry but also the high density and the low Alfv\'en speed in the streamer region, all of which are needed for the generation of type II radio emission \citep{Uchida73, Warmuth05}. 
We believe that all the above mentioned conditions, which favour the generation of type II radio bursts, were met in the studied event.

%As discussed earlier, these regions are more favourable for the production of the type II radio emission \citep[][and references therein]{Mann95book}.

\section{Summary and discussion} \label{Sec:Summary}

We present a multiwavelength analysis of the CME/flare event on September 27, 2012. The studied C3.7 flare was associated with a full-halo CME (3D speed 
\mbox{$\approx$~1300~km/s}), an EIT wave, a coronal dimming, and a WL shock. The speeds of the two type II bursts obtained employing classical method were 1500 and 1000 km/s for (HF and LF type II, respectively). 3D information on the sources of the radio emission was obtained employing the radio triangulation technique and direction-finding observations. Radio triangulation revealed the existence of two groups of type III bursts, the flare impulsive (FI) and the flare decay (FD) phase type III bursts. The FI-type IIIs had source regions close to the west CME flank, and FD-type IIIs close to the east CME flank. The LF-type II and FD-type III were found to be roughly co-spatial, appearing in the south-east quadrant of the Sun and close to the eastern flank of the CME. All the studied radio bursts originated from regions of higher density than suggested by the 1D models. The obtained density profiles crossed several different 1D models. We attributed this behaviour to the strongly non-radial propagation of the radio source.

We found that the EIT wave speed increased from 320 to 770~km/s when the wave passed through a nearby active region (i.e. the south-east direction from the source region). The modelled EIT wave dome showed a quasi-perpendicular geometry, favourable for the generation of type II radio emission, in two regions of the dome, roughly corresponding to the CME flanks. Further, we found a good correlation between the position of the LF-type II sources and the nearby streamers. This indicated that the LF-type II radio emission was generated by the shock wave/streamer interaction, similar to studies by \cite{Magdalenic14}, \cite{Zucca18}, and \cite{Mancuso19}. As the HF-type II was observed at a lower frequency than the usual metric-type II bursts \citep[above 150 MHz,][]{Klassen03, Magdalenic10, Magdalenic12}, it was probably not flare-driven. We believe that both of the type IIs are CME-driven. The difference in their starting frequency (and radio source positions) is due to the fact that they are generated in the quasi-perpendicular shock wave regions roughly corresponding to different foot-points of the same CME.

%%%%%%%%%%%%%%%%%%%%%%%%%%%%%%%%%%%%%%%%%%%%%%%%%%%%%%%%%%%%%%%%%%%%%%%%%%%%%%%%%%%%%%%%%%%
\begin{figure*}[ht]
     \centering
     \begin{subfigure}[b]{0.99\textwidth}
         \centering
         \includegraphics[width=\textwidth]{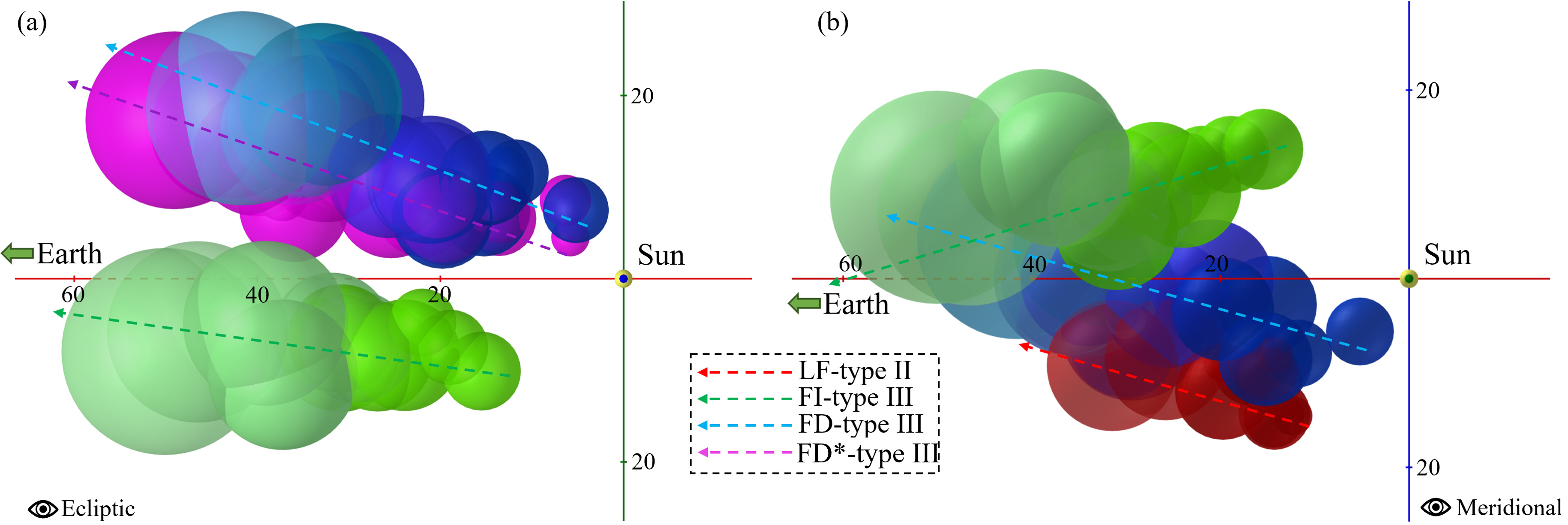}
     \end{subfigure}
     \caption{Radio triangulation results of FI-type III, FD-type III, FD*-type III, and LF-type II plotted together. The units are Solar radii ($R_{\odot}$) with the Sun at the centre ($0,0,0$). The dotted lines mark the linear-fit to the centroids of the radio sources. \textit{(a)} The results in the ecliptic plane. Propagation of all three type III sources are highly non-radial and the two FD-type III bursts seem have a similar propagation path. \textit{(b)} The results in the meridional plane. All three bursts show a propagation from high latitude towards the ecliptic plane.}
     \label{Fig:all_points_ecl_mer}
\end{figure*}
%%%%%%%%%%%%%%%%%%%%%%%%%%%%%%%%%%%%%%%%%%%%%%%%%%%%%%%%%%%%%%%%%%%%%%%%%%%%%%%%%%%%%%%%%%%%%

%%%%%%%%%%%%%%%%%%%%%%%%%%%%%%%%%%%%%%%%%%%%%%%%%%%%%%%%%%%%%%%%%%%%%%%%%%%%%%%%%%%%%%%%%%%%
\begin{figure*}[ht]
     \centering
     \begin{subfigure}[b]{0.95\textwidth}
         \centering
         \includegraphics[width=\textwidth]{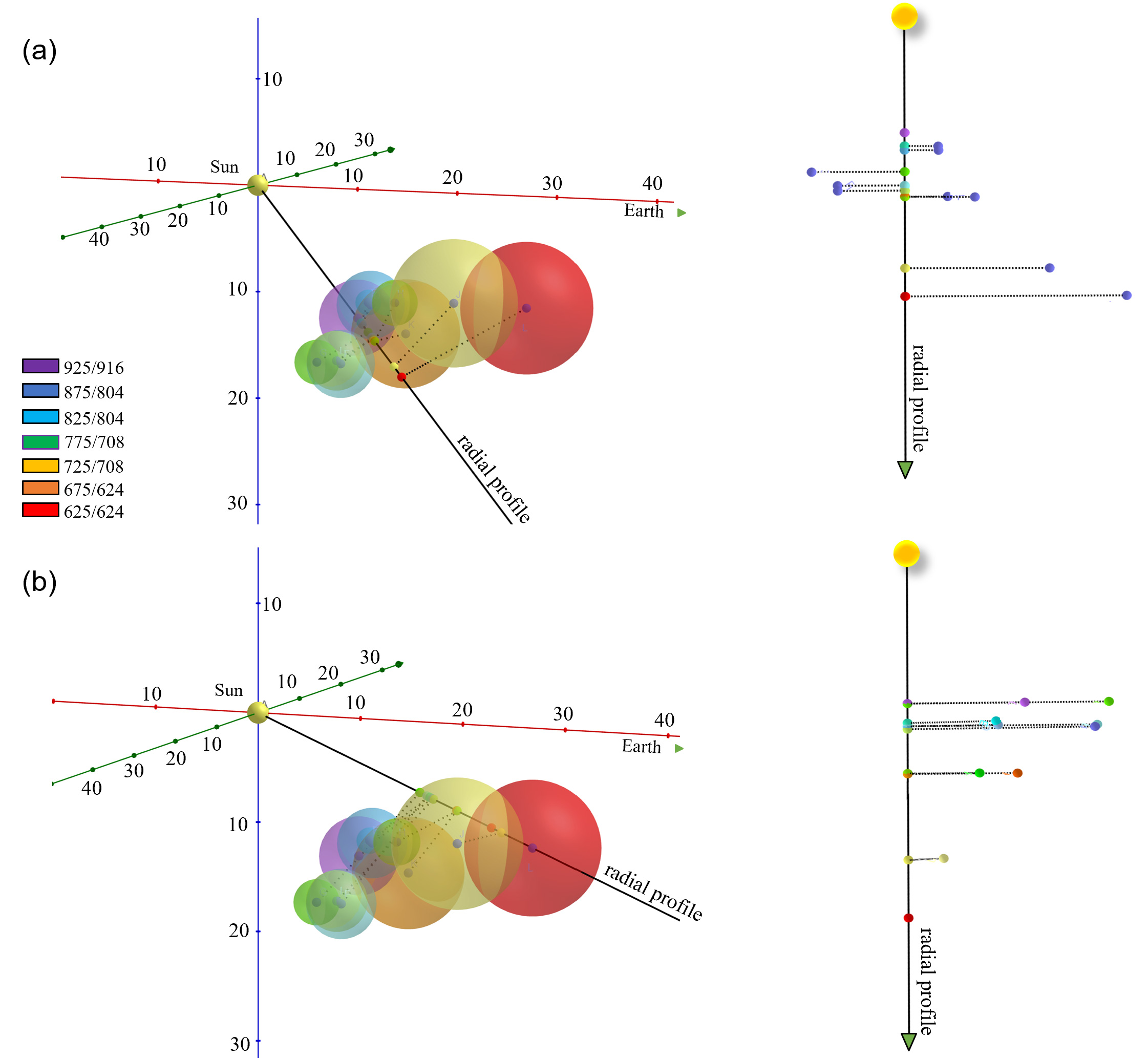}
     \end{subfigure}
     \caption{Radio triangulation results of LF-type II plotted in detail to show the effect of non-radial propagation of the emitting source. The units are Solar radii ($R_{\odot}$), and the Sun is at the center of the coordinate system. A ray is drawn from the centre of the Sun $(0,0,0)$ or $R_{\odot} = 0$ to centres of different sources. \textit{(a)} shows the result when the ray is drawn employing the highest frequency (Purple, 925/916~kHz) pair. The projection is shown in detail in the line profile adjacent to it. \textit{(a)} shows the result when the ray is drawn towards the lowest frequency (Red, 625/624~kHz) pair. The result of the orthogonal projection is shown in the profile adjacent to it.}
     \label{Fig:Projection_tpyeII}
\end{figure*}
%%%%%%%%%%%%%%%%%%%%%%%%%%%%%%%%%%%%%%%%%%%%%%%%%%%%%%%%%%%%%%%%%%%%%%%%%%%%%%%%%%%%%%%%%%%%%%%%

\subsection{Propagation of the radio emission} \label{Sec:Discussion}

During its propagation through the corona, radio emission can be modified in different ways. Two of the most frequently discussed phenomena are the scattering processes due to local density modulations along the radio emission path \citep[e.g.][]{Fokker65, Hollweg68, Riddle72, Bastian94, Arzner99}, and the non-radially propagating source of radio emission \citep[e.g.][]{Kundu65, Kai69, Nelson75, Bougeret85, Zucca18}. Due to scattering processes, the position of the radio source might be observed as shifted in comparison with its real position, and its apparent size increased \citep[see e.g.][]{Steinberg71, Kontar17b}. Different levels of the density fluctuations and their influence on scattering were recently discussed through different approaches \citep[][]{Thejappa07, Krupar18, Chrysaphi18}. We note that the majority of these studies consider radial 1D coronal electron density profiles as an input. Radio triangulation is also subject to radio-wave scattering effects, potentially inducing large distances between the wave vectors \citep{Thejappa12, Krupar16} and accordingly large source regions. As the scattering effects increase with the decrease of the observing frequency, this effect is more pronounced for the lowest direction-finding frequencies. Taking this into account, we limited our study to frequency pairs above 500~kHz.

Fig.~\ref{Fig:all_points_ecl_mer} presents the radio triangulation results in the ecliptic and meridional plane. The radio source regions correspond to the full distance between the wave vectors (see Sec.~\ref{Sec:Radio_triangulation}). The source region diameters of type III bursts at the lowest considered frequencies are about \(17 R_{\odot}\). Both, the FI-type III and the FD-type III bursts propagate from the high latitudes towards the ecliptic plan (Fig.~\ref{Fig:all_points_ecl_mer}a). The FI-type IIIs start at the northern hemisphere and FD-type IIIs start at the southern hemisphere, and they all show non-radial propagation.
Fig.~\ref{Fig:all_points_ecl_mer}b shows FI-,  FD- and FD*-type III burst (green, blue and pink spheres, respectively). We found that the propagation path of the two FD-type IIIs (separated in time by about 5 min) is almost identical, with a difference smaller than the apparent sizes of the radio sources. If the scattering effects would be significant, we would not expect that two type III bursts have the same propagation path. The solar corona is very dynamic, and scattering of the radio emission due to density fluctuations could induce significantly different radio source positions obtained for the same frequency pairs of these two bursts. As this effect is not observed in the event under study, we believe that the accuracy of the radio triangulation results is within the limits induced by the method itself. Further, the same propagation path of the subsequent type III bursts was already reported in some other studies \citep[][]{Reiner09, Klassen18, Zhang19}. We think that the scattering, that can be strongly event dependent \citep{Aurass94, Zlotnik98}, is probably not a dominant process in this event. Even if scattering induces the shift in the source positions to larger heights, it does not significantly affect the propagation direction of the radio emission and our results on the non-radial propagation of the radio emission sources.

Although non-radially propagating radio emission sources were often discussed in the 2D plane \citep{Mann03, Carley16, Zucca18}, herein we address this effect, for the first time in 3D space. The 3D positions of the LF-type II sources propagating in a strongly non-radial direction are shown in Fig.~\ref{Fig:Projection_tpyeII}a (different colours represent different frequency pairs).

To demonstrate the effects of the non-radial propagation, we 'convert' the 3D positions of the type II sources into two different 1D radial profiles (Fig.~\ref{Fig:Projection_tpyeII}). We considered the projection of the 3D sources to a radial line connecting the centre of the Sun and the radio source of the highest (Fig.~\ref{Fig:Projection_tpyeII}a), and lowest (Fig.~\ref{Fig:Projection_tpyeII}b) frequency-pairs (i.e. 925/916~kHz and 625/624~kHz, respectively). The conversion resulted in two 1D profiles with orthogonal projection of the sources. The right-hand panels of Fig.~\ref{Fig:Projection_tpyeII} show how strongly the source region propagation in 1D is different from the non-radial 3D propagation. The obtained 1D profiles also significantly differ from each other, due to differently selected radial directions. The 3D source positions and the projected positions for the same frequency-pairs are strongly different depending on the selected radial profile. The distance of the 3D source position and the projected one can be as large as $13 R_{\odot}$. In the conversion process, the 3D information was completely lost resulting in erroneous 1D profiles (Fig.~\ref{Fig:Projection_tpyeII}).

If for the studied event we would use ground based interferometric observations, the type II positions would be observed like in Fig.~\ref{Fig:Streamer_LASCO_projection}b. The strongly non-radial propagation of the radio sources would be in this case observed as an almost stationary emission. Taking all this into account, we conclude that employing 1D density profiles in the study of propagation of the radio emission needs to be done having in mind large possible errors, and could be considered only as a very rough approximation. Further, estimation of the level of scattering effects \citep[][]{Kontar17b, Chrysaphi18, Mccauley18} should also take into account the possible influence of the non-radial propagation and the projection effects. \cite{Gordovskyy19}  employed different corrections for projection effects and obtained significant changes in the estimated source heights. However, due to the lack of spatial information, the results were attributed to the scattering effects. It is probable that drawing general conclusions is difficult, as both density fluctuations and propagation direction of radio emission might strongly change from event to event.

\section{Conclusions}
\label{Sec:Conclusions}

The relationship between CME/flare events, shocks and associated type II radio bursts have been extensively discussed for several decades \citep[e.g.][and references therein]{Cairns03}. This study brings some new and important findings on the association of the radio emission and solar eruptive phenomena. We list the most important results:

\begin{itemize}
    \item [\textbullet]
    The radio triangulation studies of type III bursts have been performed earlier, but we show for the first time, that the source positions of type III bursts observed during a single eruptive event were located in significantly different locations. We found that the FI-type III bursts (observed during flare impulsive phase) originate from close to the western CME-flank region, and the FD-type III bursts (observed during the flare decay phase) originate from close to the eastern CME-flank region.
    
    \item [\textbullet]
    We found the propagation path of two subsequent type III bursts are very similar (FD- and FD*-type III, Fig.~\ref{Fig:all_points_ecl_mer}a), with differences smaller than the source region sizes (i.e. distance between two wave vectors). We did not find any significant difference in the source positions for the same frequency pairs for these two bursts, which would be expected if the scattering processes in this event would be significant. The accuracy of the radio triangulation is therefore within the limits induced by the radio triangulation method.
    
    \item [\textbullet]
    One of the two type II bursts (HF- and LF-type II) associated with studied event, the LF-type II starts at an unusually low frequency.  We found that the LF-type II was associated with the interaction of the shock wave and a streamer region. 
    Although appearing at very different parts of the CME (different flanks) both of the type II radio bursts seem to be CME-driven.
    
     \item [\textbullet]
    The radio triangulation study of the LF-type II burst provides evidence of the strongly non-radial propagation of the radio sources. Although this has been already discussed previously \citep[][and references therein]{Kai69, Bougeret85}, only the 3D information obtained in the radio triangulation allows us to  quantify the effects associated with the non-radial propagation.
    
    \item [\textbullet]
    The coronal electron densities obtained in radio triangulation study show that all radio bursts in this event are generated in the regions of higher densities than usually considered when employing 1D density models. This can be expected in particular during periods of high solar activity and at times when the global magnetic field of the Sun is very complex. Therefore, employing the 1D density models for explaining radio emission should be considered with great care, and only as a first level approximation.
    
    \item [\textbullet]
    The EIT wave, associated with eruptive event, accelerates (from 320 to 770~km/s) when passing a nearby active region, in the direction which roughly coincides with the propagation direction of the LF-type II.
    The reconstructed dome of the EIT wave indicates the existence of two main regions with quasi-perpendicular shock regimes, roughly associated with the CME-flanks. The south-west region is most probably the source region of the HF-type II, and the south-east region of the quasi-perpendicular geometry is the source region of LF-type II.
\end{itemize}

Radio triangulation is not dependent on a density model, and thus provides us a unique opportunity to study different aspects of the radio bursts and their association with the solar transients. During its propagation through the corona, the radio emission can be influenced in different ways, and this will also affect the results of the radio triangulation. Therefore, as all other observations, gonipolarimetric observations need to be treated with care, having in mind their limitations. Nevertheless, direction-finding observations provide unique information on the 3D positions of the radio emission, and can help us understand the processes of radio emission during the eruptive events in an unprecedented way.

\begin{acknowledgements}
      EIT  and  LASCO  data  have  been  used  courtesy  of  the SOHO/EIT and SOHO/LASCO consortiums, respectively. The STEREO SECCHI  data  are  produced  by  a  consortium  of  RAL(UK), NRL(USA), LMSAL(USA), GSFC(USA), MPS(Germany), CSL(Belgium), IOTA(France), and IAS(France). The Wind/Waves instrument was designed and built as a joint effort of the Paris-Meudon  Observatory,  the  University  of  Minnesota,  and the  Goddard  Space  Flight  Center,  and  the  data  are  available at the instrument Web site. We thank the radio monitoring service at LESIA (Observatoire de Paris) for providing value-added data that have been used for this study. We are grateful to the staff of the Bruny Island Radio Spectrometer for their open data policy.  The authors are grateful for useful discussions with Dr. Bojan Vr\v{s}nak, Dr. Eduard Kontar, and Dr. Milan Maksimovic, regarding the propagation of the radio emission and scattering effects. The authors are also thankful to the anonymous referee for their valuable input which helped us to significantly  improve the manuscript.
      I.C.J. was supported by a PhD grant awarded by the Royal Observatory of Belgium. C.S. acknowledges funding from the Research Foundation - Flanders (FWO, fellowship no. 1S42817N).  K.D. and A.M.V. acknowledge funding by the Austrian Space Applications Programme of the Austrian Research Promotion Agency FFG (ASAP-11 4900217  BMVIT)  and the Austrian Science Fund FWF: P24092-N16 and P27292-N20. V.K. acknowledges support by an appointment to the NASA postdoctoral program at the NASA Goddard Space Flight Centre administered by Universities Space Research Association under contract with NASA and the Czech Science Foundation grant 17-06818Y. E.K. acknowledges Finnish Centre of Excellence in Research of Sustainable Space (Academy of Finland grant number 1312390), European Research Council (ERC) under the European Union's Horizon 2020 research and innovation programme (ERC-COG 724391), and Academy of Finland project SMASH no. 310445. EUHFORIA is developed as a joint effort between the University of Helsinki and KU Leuven. The validation of solar wind and CME modelling with EUHFORIA is being performed within the BRAIN-be project CCSOM (Constraining CMEs and Shocks by Observations and Modelling throughout the inner heliosphere; www.sidc.be/ccsom/) and BRAIN-be project SWiM (Solar Wind Modeling with EUHFORIA for the new heliospheric missions). 
      
\end{acknowledgements}

%-------------------------------------------------------------------
\bibliographystyle{aa}
\bibliography{bibtex27092012}

\end{document}